\documentclass[a4paper,11pt]{article}
\pdfoutput=1 

\usepackage{jheppub} 
\usepackage{subfigure}
\usepackage{xcolor}
\usepackage[T1]{fontenc} 
\usepackage{xspace}

\newcommand{\Sherpa}{S\protect\scalebox{0.8}{HERPA}\xspace}
\newcommand{\Rambo}{R\protect\scalebox{0.8}{AMBO}\xspace}
\newcommand{\Hone}{H\protect\scalebox{0.8}{1}\xspace}
\newcommand{\Zeus}{Z\protect\scalebox{0.8}{EUS}\xspace}
\newcommand{\HERA}{H\protect\scalebox{0.8}{ERA}\xspace}

\def\GeV{\ifmmode {\mathrm{\ Ge\kern -0.1em V}}\else \textrm{Ge\kern -0.1em V}\fi}%
\def\GeV{\ifmmode {\mathrm{\ Ge\kern -0.1em V}}\else \textrm{Ge\kern -0.1em V}\fi}%

\title{\boldmath Large Effects from Small QCD Instantons: \\  Making Soft Bombs at Hadron Colliders}

\author[a]{Valentin V. Khoze,}
\author[a]{Frank Krauss}
\author[b,c]{and Matthias Schott}

\affiliation[a]{Institute for Particle Physics Phenomenology, Durham University, UK}
\affiliation[b]{University College London, London, UK}
\affiliation[c]{Johannes Gutenberg-University, Mainz, Germany}

\abstract{
It is a common belief that the last missing piece of the Standard Model of particles physics was found with the discovery of the Higgs boson at the Large Hadron Collider. However, there remains a major prediction of quantum tunnelling processes mediated by instanton solutions in the Yang-Mills theory, that is still untested in the Standard Model. The direct experimental observation of instanton-induced processes, which are a consequence of the non-trivial vacuum structure of the Standard Model and of quantum tunnelling in QFT, would be a major breakthrough in modern particle physics. 
In this paper, we present for the first time a full calculation of QCD instanton-induced processes in proton-proton collisions accounting for quantum corrections due to both initial and final state gluon interactions, 
a first implementation in an MC event generator as well as a basic strategy how to observe these effects experimentally.
}

\begin{document}
\preprint{IPPP/19/85}

\maketitle
\flushbottom
\newpage


\section{\label{Sec:Intro}Introduction}
\medskip

In the last decades, the Standard Model of Particle Physics (SM) evolved to the most precise theory in terms of fundamental interactions of the elementary constituents of matter. With the discovery of the Higgs Boson in 2012, the last missing predicted particle was found \cite{Aad:2012tfa, Chatrchyan:2012xdj}. With its mass value of 125 GeV, we finally have a theory which could in principle remain valid up to the Planck scale and could describe all interactions except gravity to timescales down to $10^{-43}$\,s after the Big Bang. 

Despite its great success, there remain open questions: The SM cannot account for the dark matter content of the universe and inhibits several fine-tuning problems, e.g.\ it does not explain why the Higgs boson mass is so much smaller than the Planck mass or why QCD does not break CP-symmetry. 
It also does not explain the observed matter-antimatter asymmetry of the universe, which must have been present shortly after the Big Bang, and this remains one of the major outstanding questions in modern physics.

One of the most popular approaches to address this issue is Leptogenesis \cite{Fukugita:1986hr}. In this framework the SM is extended with ultra-heavy right-handed Majorana neutrinos, which then decay into ordinary SM leptons and the Higgs, thus generating the SM lepton number asymmetry.
 The generated asymmetry of leptons is then converted to a baryonic asymmetry through instanton-like non-perturbative processes in the Weinberg-Salam theory. The fact that instantons violate baryon plus lepton number in the electroweak sector of the Standard Model is well-known
 \cite{tHooft:1976rip}, however in normal circumstances the effect is exponentially suppressed by the instanton action, $R_{B+L} \propto e^{-S_I} = e^{-4\pi/\alpha_w} \lll 1$, often referred to as the 't Hooft suppression factor. But in the early universe, at temperatures above the electro-weak phase transition,
 the exponential suppression disappears along with the barrier separating the SM vacua with different baryon and lepton numbers  \cite{Kuzmin:1985mm},
 thus enabling the Thermal Leptogenesis mechanism \cite{Fukugita:1986hr} to complete and generate the observed matter-anti-matter asymmetry of the universe.
 
The non-Abelian nature of Yang-Mills theories implies a non-trivial vacuum structure~\cite{Callan:1976je,Jackiw:1976pf}. 
While ordinary perturbation theory works well for most processes of the SM, the instanton processes correspond to quantum tunnelling between different vacuum sectors, and cannot be described with the usual perturbative approach. Instantons \cite{Belavin:1975fg} are manifestly non-perturtbative semiclassical contributions to the path integral; they are directly related to anomalous Ward identities \cite{Adler:1969gk,Bell:1969ts,Bardeen:1969md} and lead to the violation of baryon plus lepton number (B+L) in the electroweak theory  as well as to chirality violation in QCD \cite{tHooft:1976snw, tHooft:1976rip}. 
Instanton-like topological fluctuations of the gauge fields have been argued to play an important role in various long-distance aspects of QCD, and as such provide a possible solution to the axial $U(1)$ problem \cite{tHooft:1986ooh}
and generate the QCD axion potential \cite{Weinberg:1977ma,Wilczek:1977pj}.
It was also demonstrated in \cite{Dorey:1996hu,Nekrasov:2002qd,Dorey:2002ik}
that instanton contributions are fully calculable and play a crucial role in the non-perturbative dynamics of supersymmetric gauge theories 
\cite{Seiberg:1994rs,Affleck:1983mk}
and in the AdS/CFT correspondence~\cite{Maldacena:1997re,Banks:1998nr,Dorey:1999pd}.
But at the same time, even though instanton processes are a core prediction of the Standard Model, they have never been experimentally observed.

\medskip

The main focus of this paper is QCD instantons and their contributions to high-energy scattering processes at hadron colliders in general and in particular 
at the LHC. Quantum corrections to the leading-order instanton contributions are critically important in QCD as they are known to contribute to the exponent of the instanton cross-section. 
For the first time we will include the quantum effects arising from both: the final state and the initial state interactions in the instanton background.
We will achieve this by combining the methods pioneered 
in \cite{Khoze:1990bm,Khoze:1991mx} and \cite{Mueller:1990qa,Mueller:1991fa} for computing quantum effects due to the final-state rescatterings 
and the initial state interactions respectively.

\medskip

The question whether manifestations of tunnelling processes in QFT can be directly observed in high-energy experiments 
was already raised in the 1990s in the context of the electro-weak theory~\cite{Ringwald:1989ee,Espinosa:1989qn,Aoyama:1986ej,McLerran:1989ab,Zakharov:1990dj,Yaffe:1990iy,Khlebnikov:1990ue,Khoze:1990bm,Mueller:1990qa,Mueller:1990ed,Khoze:1991mx,Mueller:1991fa}.
Studies of collider phenomenology of electro-weak instantons were carried out in~\cite{Gibbs:1994cw} and more recently 
in~\cite{Ringwald:2018gpv,Papaefstathiou:2019djz}. 

An obvious way to reduce the semiclassical 't Hooft suppression instanton factor is to consider QCD instantons since 
the suppression is exponential, $e^{-4\pi/\alpha}$, 
and the strong coupling coupling constant is $\alpha_s \gg \alpha_w$.
Most of QCD instanton-induced hard-scattering processes studied in the literature were specific to deep-inelastic 
scattering (DIS)~\cite{Balitsky:1993jd,Ringwald:1994kr,Moch:1996bs,Ringwald:1998ek}. In this case, the instanton process kinematics is characterised by two
scales:  the CoM energy $\sqrt{s}$, as well as the deep inelastic momentum scale $Q$. The existence of the latter scale representing
the virtuality of one of the incoming particles in the collision,
was essential for obtaining infrared safe instanton contributions in the DIS settings.
It introduced a factor of $e^{-Q\rho}$ in the amplitude of the process \cite{Khoze:1991mx,Balitsky:1992vs}
and that enabled an effective cut-off of the integrations over the large instanton sizes 
$\rho$ in this approach. 

\medskip
The \Hone and \Zeus Collaborations have searched for QCD instantons at the \HERA collider~\cite{Carli:1997tw, Carli:1998zf, Adloff:2002ph, Chekanov:2003ww, H1:2016jnv}.  The observables used to discriminate the instanton-induced contribution from that of perturbative DIS processes, are based on the hadronic final state objects and on a selection of charged particles. The searches were therefore based on assuming an isotropic decay in the centre-of-mass frame into ${\cal{O}}(10)$ partons 
 plus, potentially, one highly energetic jet in the forward region, where the virtuality of the incident photon $Q$ sets the scale for the process and the instanton size.  With all light quark flavours equally present in the final state (flavour democracy), several strange mesons and baryons such as $K^\pm$ and $\Lambda$'s were also expected.  A multivariate discrimination technique was employed by \Hone to increase the sensitivity to instanton processes, leading to the strongest upper limits. They range between 1.5 pb and 6 pb, at 95\% confidence level, depending on the chosen kinematic domain. While this result challenges the predictions based on the lattice data of Ref. \cite{Smith:1998wt}, it is fully compatible \cite{Schremp2016p} with the expectations based on the lattice data of Ref. \cite{Hasenfratz:1998qk}, see also Ref.~\cite{Athenodorou:2018jwu}.

On the other hand, for generic scattering processes at hadron colliders -- the settings relevant to  this paper -- we do not have a second independent 
kinematic scale, such as the DIS highly virtual momentum scale $Q$. In particular, 
both incoming partons are on their mass-shell (i.e.\ have no large virtualities) and we do not want to introduce any unnatural bias into the final state, for example by demanding a high-mass photon or gauge boson that decays into leptons. The dominant instanton-induced process has, as we will see, 
an unbiased isotropic multi-particle final state. As a consequence, QCD instanton-induced scattering processes produce {\it soft bombs} -- very high-multiplicity
spherically symmetric distributions of relatively soft particles.  The phenomenology of such events, usually associated with Beyond the Standard Model effects,  was first investigated in~\cite{Knapen:2016hky}, but in our case the soft bombs will be fully Standard Model-made: they will be generated by the QCD instantons.

In our approach, only small instantons contribute to the scattering processes in QCD. The potentially problematic contributions of instantons with large size
are automatically cut-off by the inclusion of quantum effects due to interactions of the hard initial states that generate the factor
 $e^{- \alpha_s \, \rho^2 s' \log s'}$, as we will explain in Section~\ref{Sec:Calculation}.
This provides a dynamical solution to the well-known problem of IR divergences arising from instantons of large scale-sizes in QCD. The main point is that 
these quantum effects break the apparent scale invariance of the classical Yang-Mills theory by lifting the classically flat instanton size mode and suppressing all 
but small instantons with sizes $\rho \lesssim (10-30) / \sqrt{s'}$. 

The fact that the characteristic instanton size in QCD is inversely proportional to the centre-of-mass (CoM) energy of two colliding partons $\sqrt{s'}$, 
and hence becomes smaller and smaller as
one increases $\sqrt{s'}$, allows to circumvent the general believe that `one cannot make a fish at a hadron  collider'~\cite{Mattis:1991bj}.
The two initial hard partons can be thought of as wave-packets of size $d\sim 1/(2 \sqrt{s'})$. This makes it very difficult to produce an electro-weak 
sphaleron which has the spatial extend of $1/M_W$ which is much greater than the inverse energy of the order of the sphaleron mass. 
Based on this intuitive picture it was pointed out in~\cite{Banks:1990zb}
 that in the electro-weak theory an instanton-induced process describing a scattering of two hard initial particles would remain
exponentially suppressed at any energies, even much above the sphaleron mass. These expectations were confirmed with a 
numerical evaluation of classical scattering rates at energies {\it above} the sphaleron barrier in 
Refs.~\cite{Bezrukov:2003er,Bezrukov:2003qm}.
In QCD, on the other hand, our results show that instantons are relatively small and the corresponding effective QCD sphaleron size 
in fact falls with the increasing $\sqrt{s'}$, thus avoiding any additional excessive exponential suppression of the scattering rates.

Finally it is important to point out that in our case the potentially observable instanton cross-sections do not require a very substantial compensation of the
original 't Hooft suppression factor in the exponent. The combination of large pre-factors in front of the exponent (that we compute) and the fact that the 
QCD coupling at the instanton scale $\rho$ is in the range
$0.1 \lesssim \alpha_s \lesssim 0.4 $
 (that is $\alpha_s $ far not as small as in the electro-weak case) makes it possible to achieve sufficiently large cross-sections in the regime
where the 't Hooft suppression in the exponent is reduced by only $\sim 20-30\%$.\footnote{This can be inferred from the plot of the normalised instanton-anti-instanton action 
$-S(\chi)$ in Fig.~\ref{fig:saddle}. Full 't Hooft suppression would correspond to $S=1$. }
This fact improves the theoretical robustness of the calculation by reducing any potential impact of even higher-order quantum corrections to our result.  It also justifies neglecting higher-order multi-instanton--anti-instanton configurations,
that were considered in Refs.~\cite{Zakharov:1990xt,Maggiore:1991vi,Veneziano:1992rp}
and were argued to set a limit on the applicability of the instanton calculation in the regime
where the 't Hooft instanton suppression is reduced by $\gtrsim 50\%$.

\medskip
\section{\label{Sec:Calculation}Instanton Cross-section Calculation}

 \subsection{QCD instanton preliminaries}
 \label{sec:prelim}

Instanton~\cite{Belavin:1975fg}
is the solution of the classical equations of motion in Euclidean spacetime; for QCD the instanton field configuration 
involves the gluon component $A_\mu^{\rm inst}$ as well as the fermion components -- the fermion zero modes  $\psi^{(0)}$.
The QCD instanton of topological charge $Q=1$ has two fermion zero modes 
for each of the $f=1,\ldots, N_f$ light quark flavours; they correspond to the Weyl fermions
$\bar{q}_{Lf}$ and $q_{Rf}$.
Light
flavours are those that can be resolved by the instanton of size $\rho$, that is with their masses $m_f \le 1/\rho$.\footnote{The instanton
size $\rho$ will ultimately be set by the energy (or other relevant kinematical variables) of the scattering  process, 
as will become clear below.}
In our notation the chiral fermions $\bar{q}_{L}$ and
 $q_{R}$ belong to the same irreducible representation of the Lorentz group, while the opposite chirality fermions 
 $q_{L}$ and $\bar{q}_{R}$ belong to the other irreducible representation. Fermion mass terms are of the form $m\, \bar{q}_{L} q_{R} + h.c.$

We will consider the instanton-dominated QCD process with two gluons in the initial state,
\begin{equation}
 g + g \, \to\, n_g \times g \,+\, \sum_{f=1}^{N_f} (q_{Rf} +\bar{q}_{Lf})\,.
 \label{eq:inst1}
 \end{equation} 
Note that the number of gluons $n_g$ in the final state is not fixed and can become large 
 even for the leading-order instanton effect (i.e.\ at leading order in instanton perturbation theory).
 On the other hand, the fermionic content of the reaction \eqref{eq:inst1} is fixed.
 The process \eqref{eq:inst1} is written for the instanton of topological charge $Q=1$, and as the result it contains precisely
 one right-handed quark  and one anti-particle of the left-handed quark for each light flavour in the final state. No fermions of opposite 
 chirality,
  i.e.\ no left-handed quarks and anti-right-handed quarks appear on the r.h.s.\ of \eqref{eq:inst1}; this being the consequence of the 
  fact that one-instanton fermion zero modes exist only for  $\bar{q}_{L}$ and $q_R$, as dictated by the
  Atyiah-Singer index theorem for the Dirac operator in the instanton background. 
 This fermion counting~\cite{tHooft:1976rip}  is also in agreement with the Adler-Bell-Jackiw anomaly.
 
 There are precisely $N_f$ of $\bar{q}_{L} q_{R}$
 pairs. We will see that in the kinematic regime relevant to our applications the condition $\rho^{-1} \gtrsim m_f$ restricts the 
 number of flavours that are counted as light to $N_f=4$ and $N_f=5$.
 The analogous to \eqref{eq:inst1} process that is induced by an 
 anti-instanton configuration, is obtained by interchanging the right-handed and the left-handed chirality labels of the fermions.
 
 We can also have quark-initiated instanton processes; they are
 obtained from \eqref{eq:inst1} by inverting two of the outgoing fermion legs in the final state
 into incoming anti-fermions in the initial state,
 giving for example,
 \begin{eqnarray}
 u_L +\bar{u}_R &\to &  n_g \times g  +\sum_{f=1}^{N_f-1} (q_{Rf} +\bar{q}_{Lf}) \,,  \label{eq:inst2a} \\
u_L + d_L &\to &   n_g \times g\,+\, u_R + d_R+ \sum_{f=1}^{N_f-2} (q_{Rf} +\bar{q}_{Lf})  \,.
 \label{eq:inst2b}  
\end{eqnarray}
Instanton contributions to all such $2 \to many$ processes \eqref{eq:inst1}-\eqref{eq:inst2b} are computed in the semiclassical approach
 by expanding
the path integral expression for the corresponding scattering amplitude around the instanton and integrating over the instanton 
collective coordinates as well as over all field fluctuations 
around the instanton~\cite{tHooft:1976snw}. 

\medskip
\noindent From now on we will concentrate on the process \eqref{eq:inst1} with two gluons in the initial state.
Quark-initiated processes can be evaluated analogously, giving partonic cross-sections of a similar order of magnitude in the semiclassical 
approximation. It is however the gluon-initiated process \eqref{eq:inst1}, that will give the dominant contribution to the hadronic instanton cross-section
thanks to large contributions of gluon parton distribution functions in the low-$x$ region.

At the leading order in the semiclassical expansion around the instanton, the scattering amplitude describing the $2 \to n_g+2N_f$ process 
\eqref{eq:inst1} is obtained by: 
\begin{enumerate}
\item Plugging the instanton solution,
\begin{equation}
A_\mu = A_\mu^{\rm inst} (x)\,, \quad
\bar{q}_{Lf} = \psi^{(0)}(x) \,, \quad q_{Rf} = \psi^{(0)}(x) \,,
\end{equation}
into external legs of the corresponding Green's function, so that it reads,
\begin{eqnarray}
 &&G_{n_g+2+2N_f} \,(x_1,\ldots, x_{n_g+2},y_1,\ldots\,y_{2N_f})\,= 
 \label{eq:Gn} \\
&& \int DA_\mu [Dq D\bar{q}]^{N_f} \,\, A^{\rm inst}_{\mu_1}(x_1)\ldots A^{\rm inst}_{\mu_{n_g+2}}(x_{n_g+2})\,
\psi^{(0)}(y_1)\ldots\psi^{(0)}(y_{2N_f}) \,\,e^{-S_E},
\nonumber
\end{eqnarray}
\item  Fourier transforming \eqref{eq:Gn} to the momentum space to obtain $\tilde{G} (p_1,p_2; k_1,\ldots, k_{n_g+2N_f})$, where
$p_i$  ($k_j$) are the momenta of the incoming (outgoing) particles,
\item Taking all momenta on-shell and performing
 the LSZ reduction for all external legs of  the Green's function $\tilde{G}$.
 \end{enumerate}

The outcome of this procedure is that the instanton contribution to the $n$-point amplitude at the leading order is recast as an
effective $n$-point vertex involving $n_g+2$ gluons and $2N_f$ quarks,
\begin{equation}
{\cal A}_{\, 2\to\, n_g+2N_f} \,\sim\,  \int  d^4 x_0 \, d\rho\, D(\rho) \,
e^{-S_I}\,
\Bigl[ \prod_{i=1}^{n_g+2} A_{\,{\rm LSZ}}^{a_i\, \rm inst}(p_i,\lambda_i)\Bigr]  \, 
\Bigl[ \prod_{j=1}^{2N_f} \psi_{\,{\rm LSZ}}^{(0)}(p_j,\lambda_j)\Bigr]\,. 
 \label{eq:Gn_LO}
\end{equation}
Here $D(\rho)$ is the instanton density, $S_I$ is the instanton action, and the field insertions are given by the LSZ-amputated instanton
solutions for gluons (see Eq.~\eqref{eq:Inst_LSZA} below) and similarly for fermions.
Because of the fully factorised structure
of the field insertions in the leading order instanton expression \eqref{eq:Gn_LO},
 there are no correlations between the momenta of the external 
legs, apart from the usual momentum conservation constraint. Emission of individual particles in the final state is independent from one another apart from the usual conservation laws.
Hence in the CoM frame, the instanton vertex  \eqref{eq:Gn_LO} describes the scattering process into 
a {\it spherically symmetric multi-particle final state}.

The instanton production cross-section $\hat\sigma$ for the process \eqref{eq:inst1}\footnote{Hat in $\hat\sigma$ 
indicates that it is a partonic cross-section.}
 can then be obtained in the usual way
by squaring the scattering amplitude and integrating over the $(n_g+2N_f)$-particle phase space including the relevant symmetry factors.
This program was developed and implemented in the classic high-energy instanton
papers \cite{Ringwald:1989ee,Espinosa:1989qn,McLerran:1989ab,Zakharov:1990dj} (for reviews see \cite{Mattis:1991bj,Voloshin:1994yp}) in the context of the 
electroweak theory for $(B+L)$-violating processes.

\subsection{The optical theorem on the instanton--anti-instanton configuration}
\label{sec:iibar}

An equivalent and arguably more direct way to obtain a total parton-level instanton cross-section $\hat\sigma_{\rm tot}^{\rm inst}$
for the process $gg \to X$, is to use the optical theorem, and compute an imaginary part of the $2\to2$ forward elastic scattering
amplitude, ${\cal A}^{{I\bar{I}}}_4 (p_1,p_2,-p_1,-p_2)$,
in the background of an instanton--anti-instanton configuration, following the approach 
initiated in~\cite{Khoze:1990bm,Khoze:1991mx},
\begin{eqnarray}
\hat\sigma_{\rm tot}^{\rm (cl)\,inst} &=&  \frac{1}{s'}\, {\rm Im} \, {\cal A}^{{I\bar{I}}}_4 (p_1,p_2,-p_1,-p_2) \nonumber  \\
&\simeq& \frac{1}{s'}\, {\rm Im}  \int_0^\infty d\rho \int_0^\infty d\bar{\rho}\,\int d^4R \,\int d\Omega\,\,
D(\rho) D(\bar{\rho}) \,\,
e^{-S_{I\bar{I}}}\,\, {\cal K}_{\rm ferm}
\times \nonumber  \\
&& \, A^{\rm inst}_{LSZ}(p_1)\,A^{\rm inst}_{LSZ}(p_2)\,A^{\overline{\rm inst}}_{LSZ}(-p_1)\,A^{\overline{\rm inst}}_{LSZ}(-p_2)
\,,
\label{eq:op_th}
\end{eqnarray}
Below we explain this formula in detail.

The integrals are over all collective coordinates of the instanton--anti-instanton configuration:
 $\rho$ and $\bar{\rho}$ are the instanton and anti-instanton sizes;
$R_\mu$ is the separation between the $I$ and $\bar{I}$  positions in the Euclidean space
and, finally, $\Omega$ is the $3\times 3$ matrix that specifies the relative ${I\bar{I}}$ orientation in the $SU(3)$ colour space.

The instanton density appearing in the integration measure in \eqref{eq:op_th} is given by the 1-loop expression \cite{tHooft:1976snw},
\begin{equation}
D(\rho,\mu_r)\,=\, 
\kappa \, \frac{1}{\rho^5} \left( \frac{2\pi}{\alpha_s(\mu_r)}\right)^6\, (\rho \mu_r)^{b_0}\,
\label{eq:Imeasure}
\end{equation}
where
$\mu_r$ is the renormalization scale, $b_0=11-2/3 N_f$,  and the constant 
$\kappa$ (computed in the $\overline{\rm MS}$ scheme) is,
 \begin{equation}
 \kappa\,\approx\, 0.0025 \, e^{0.291746 N_f} \,, \quad {\rm so\, that} \quad
  \kappa_{N_f=4} \,\approx\, 0.008\,, \quad
   \kappa_{N_f=5} \,\approx\, 0.01\,.
 \end{equation}

The exponential factor $e^{-S_{I\bar{I}}}$ in \eqref{eq:op_th} is the semiclassical suppression
factor of the process by the action  of the instanton--anti-instanton configuration,
\begin{equation}
S_{I\bar{I}} \,=\, S_I \,+\, S_{\bar{I}}\,+\, U_{\rm int}(\rho,\bar{\rho},R,\Omega)  \,,
\end{equation}
where $S_I = S_{\bar{I}} =\frac{2\pi}{\alpha_s(\mu_r)}$ is the action of a single (anti)-instanton, and 
$U_{\rm int}(\rho,\bar{\rho},R,\Omega)$ is the interaction potential between the instanton and the anti-instanton.
The interaction potential can be repulsive or attractive, depending on the choice of the relative orientation $\Omega$. 
In the steepest-descent approximation, the integrand in \eqref{eq:op_th} will be dominated by the saddle-point solution
that extremises the function in the exponent. This corresponds to the maximally attractive interaction channel, i.e.\
the value of $\Omega$ for which $- U_{\rm int}(\rho,\bar{\rho},R,\Omega)$ is maximal, or equivalently,
the action $S_{I\bar{I}} $ is minimal (for fixed $R$ and $\rho,\bar{\rho}$).

The general expression for the action as the function of $R, \rho,\bar{\rho}$ was  
computed in \cite{Khoze:1991mx} 
using the form of the 
instanton--anti-instanton valley configuration \cite{Balitsky:1986qn,Yung:1987zp,Khoze:1991sa} dictated by the conformal invariance of classical Yang-Mills theory.
For the maximally attractive relative orientation, the action takes the form \cite{Khoze:1991mx},
\begin{eqnarray}
S_{I\bar{I}}(\rho, \bar{\rho},R) &=&
\frac{4\pi}{\alpha_s(\mu_r)} \,\hat{\cal S}  \,,
\label{eq:Shat} \\
\nonumber\\
\hat{\cal S} &=& 3\left(\frac{6z^2-14}{(z-1/z)^2}\,-\, \frac{17}{3}\,-\, \log(z) \left(
 \frac{(z-5/z)(z+1/z)^2}{(z-1/z)^3}-1\right)\right)\,,
 \label{eq:Szdef}
\end{eqnarray}
where $z$ is a conformal ratio of the instanton collective coordinates,
\begin{equation}
z\,=\, \frac{R^2+\rho^2+\bar{\rho}^2+\sqrt{(R^2+\rho^2+\bar{\rho}^2)^2-4\rho^2\bar{\rho}^2}}{2\rho\bar{\rho}}\,.
\label{eq:zdef}
\end{equation}
Thus, the expression for the instanton--anti-instanton action \eqref{eq:Szdef} is a function of a single argument $z$ that is obtained 
from the instanton--anti-instanton separation $R$, $R^2=R_\mu R_\mu = R_0^2 + \vec{R}^2$, and the scale sizes $\rho$ and $\bar{\rho}$, as defined in \eqref{eq:zdef}.  

In the limit of large separation between the instanton centres, $R/\rho,\, R/\bar{\rho} \to \infty$, the conformal ration $z\to R^2/\rho\bar\rho \to \infty$, and the
function $\hat{\cal S}$ in \eqref{eq:Szdef} goes to 1. 

One can also verify that in the opposite limit of the vanishing separations $R/\rho,\, R/\bar{\rho} \to 0$
that corresponds to $z\to 1$, the expression on the r.h.s. of \eqref{eq:Szdef} for the normalised action $\hat{\cal S}(z)$ goes to zero,
\begin{equation}
\lim_{z\to 1} \hat{\cal S} \,=\, \lim_{z\to 1}\, \frac{2}{5}(z-1)^2 \,+\, {\cal O}  (z-1)^3 \,= \, 0.
\nonumber
\end{equation}

Motivated by the symmetry between the instanton and the anti-instanton, and to better visualise the dependence of the instanton--anti-instanton action 
on instanton collective coordinates, we can consider a slice $\rho=\bar{\rho}$ and introduce a new dimensionless variable 
\begin{equation}
\chi=R/\rho\,,
\end{equation}
to characterise the relative ${I\bar{I}}$ separation. The instanton--anti-instanton action is then a function of $\chi$,
\begin{equation} 
S_{I\bar{I}}(\rho, \bar{\rho},R) \,=\, 
\frac{4\pi}{\alpha_s(\mu_r)} {\cal S}(\chi)\,,
\label{eq:Schi}
\end{equation}
where
\begin{equation}
\hat{\cal S}(\chi)\,=\, \hat{\cal S}(z(\chi)) \,, \quad {\rm and} \quad
z\,=\, \frac{1}{2}\left(\chi^2+\chi\sqrt{\chi^2+4} +2\right)\,.
\label{eq:zchi}
\end{equation}
At large separations, $\chi \gg 1$, the expression \eqref{eq:Szdef} for the instanton--anti-instanton action simplifies and reduces to the 
well-known in the early instanton literature result,
\begin{equation}
{\cal S}(\chi)\,\simeq \,  
1\,-\, 6/\chi^4\,+\, 24/\chi^6 \,+\, \ldots
\label{eq:zchi2}
\end{equation}
The first term in the $I\bar{I}$ interaction, $-6/\chi^4$ effectively takes into account the effects
of the $n_g$ final  state gluons in the amplitude \eqref{eq:Gn}  \cite{Zakharov:1990dj,Khoze:1990bm}. The next term,
$24/\chi^6 $, computed originally in \cite{Khoze:1990bm}, accounts for the leading-order interactions between the final state gluons.
These results were successfully tested against the direct calculation of the
interactions between the final state gluons, the so-called final-final state interactions \cite{Mueller:1991fa,Khlebnikov:1990ue}.

\begin{figure}[]
\begin{center}
\begin{tabular}{cc}
\hspace{-.4cm}
\includegraphics[width=0.5\textwidth]{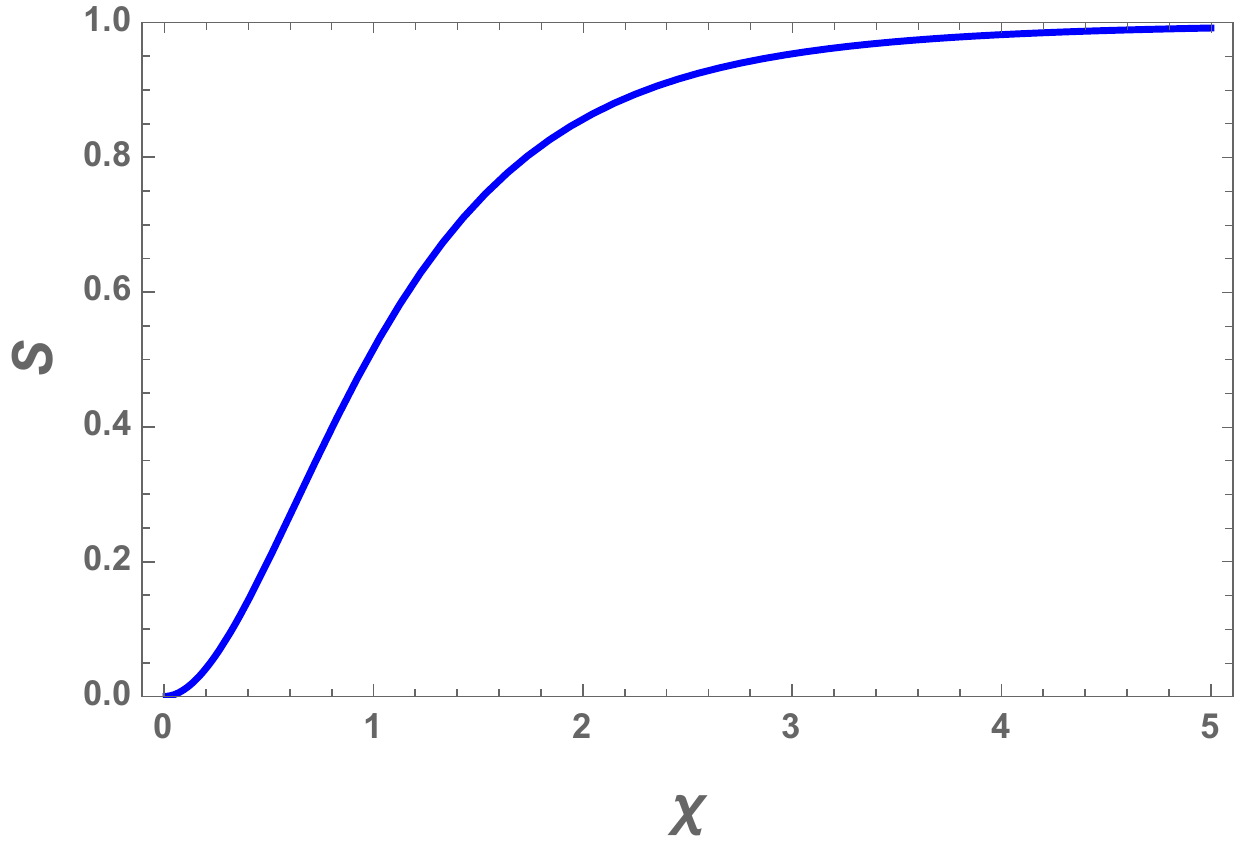}
&
\includegraphics[width=0.5\textwidth]{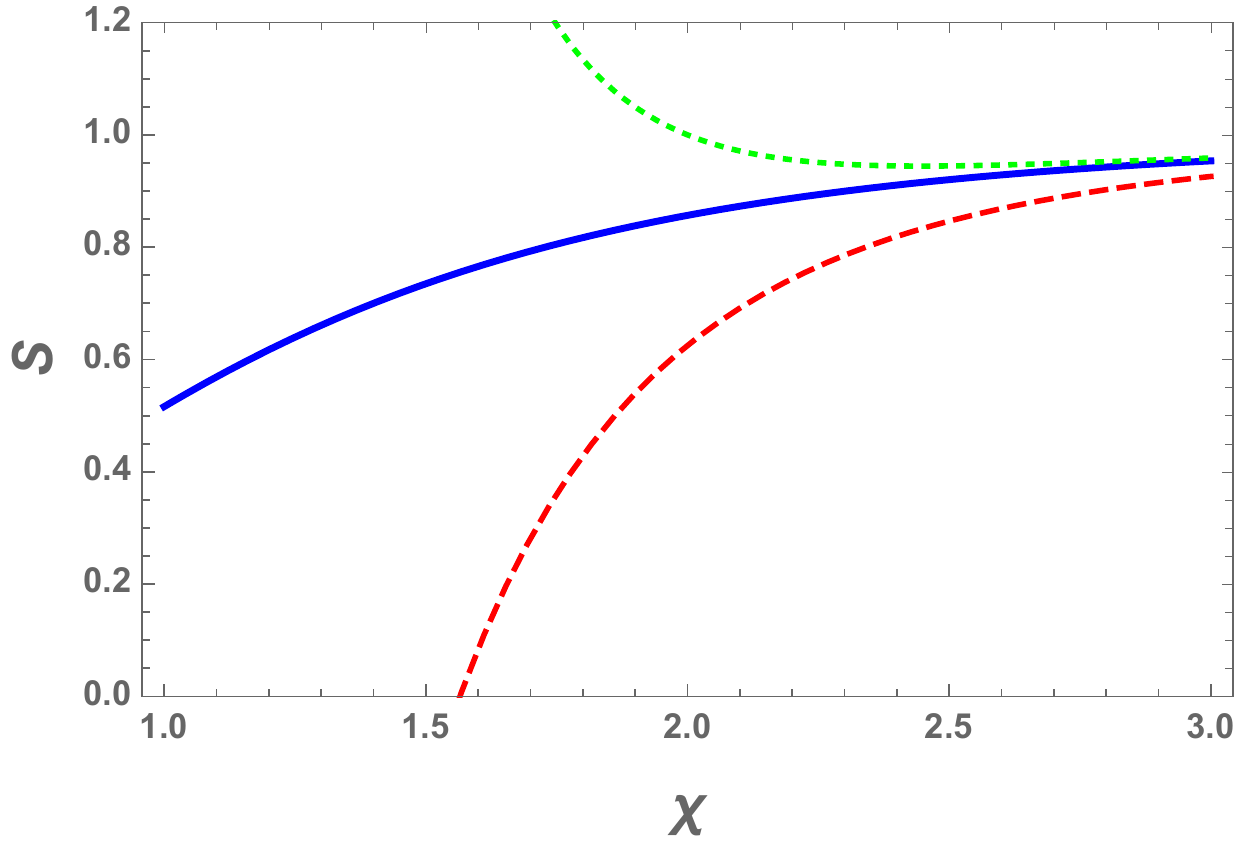}
\\
\end{tabular}
\end{center}
\vskip-.4cm
\caption{
The action \eqref{eq:Szdef} of the instanton--anti-instanton configuration as the function of $\chi=R/\rho$ (solid line).
${\cal S}(\chi)$ approaches one
at $\chi\to \infty$ where the interaction potential vanishes, and ${\cal S}\to 0$ at $\chi\to 0$ where the
 instanton and the anti-instanton 
mutually annihilate. 
The plot on the right also shows
the leading-order (dashed line) and the next-to-leading-order (dotted line) approximations in \eqref{eq:zchi2}. The regime
of interest to us is $1.5<\chi<1.7$, where we have to use the complete action (solid line).}
\label{fig:S}
\end{figure}

In the kinematic regime we study in this paper, 
the value of the $\chi$ variable at the saddle-point will turn out to be in the interval $1.5<\chi<1.7$ 
which requires the use of the complete expression for the $I\bar{I}$ action given in \eqref{eq:Szdef}, \eqref{eq:zchi}.
The expression we use for ${\cal S}(\chi)$ is plotted in Fig.~\ref{fig:S}.

\begin{figure}[]
\begin{center}
\begin{tabular}{cc}
\hspace{-.4cm}
\includegraphics[width=0.5\textwidth]{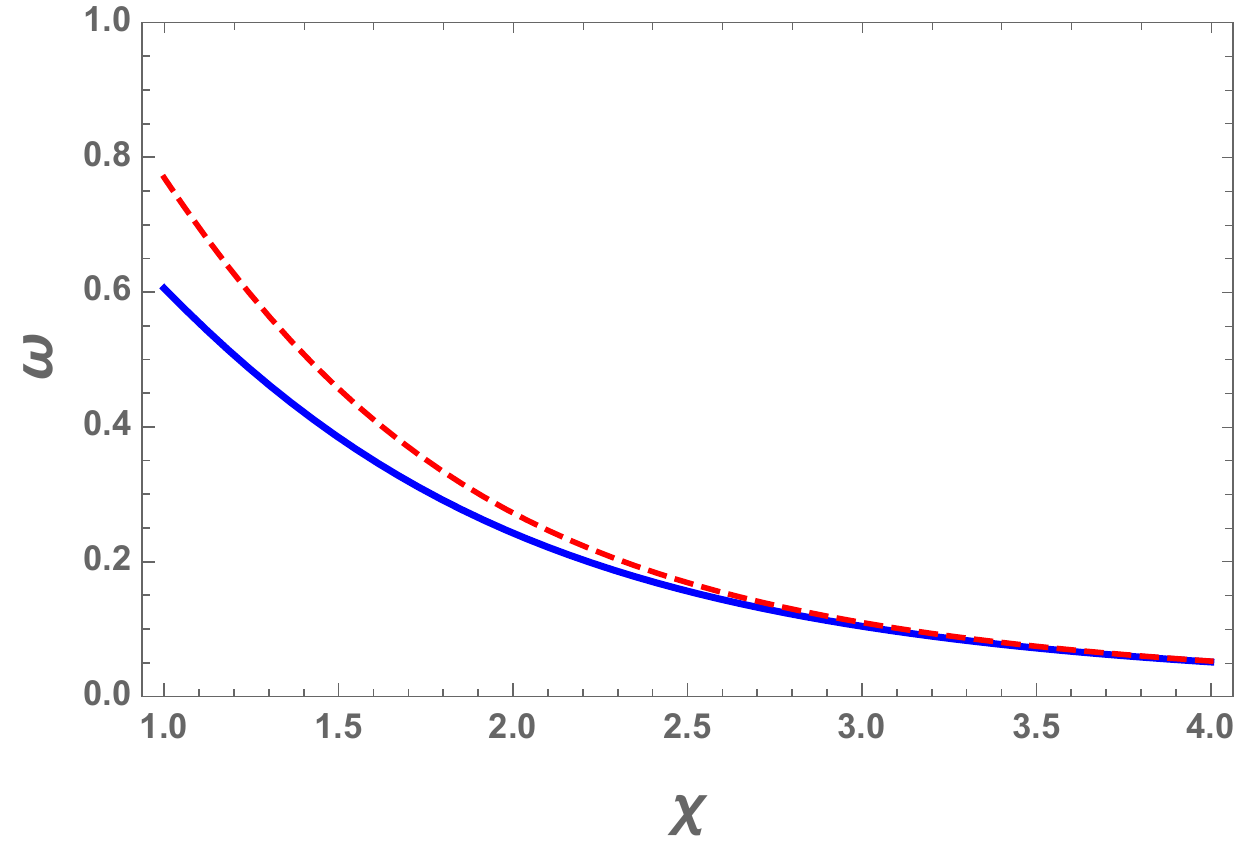}
&
\includegraphics[width=0.5\textwidth]{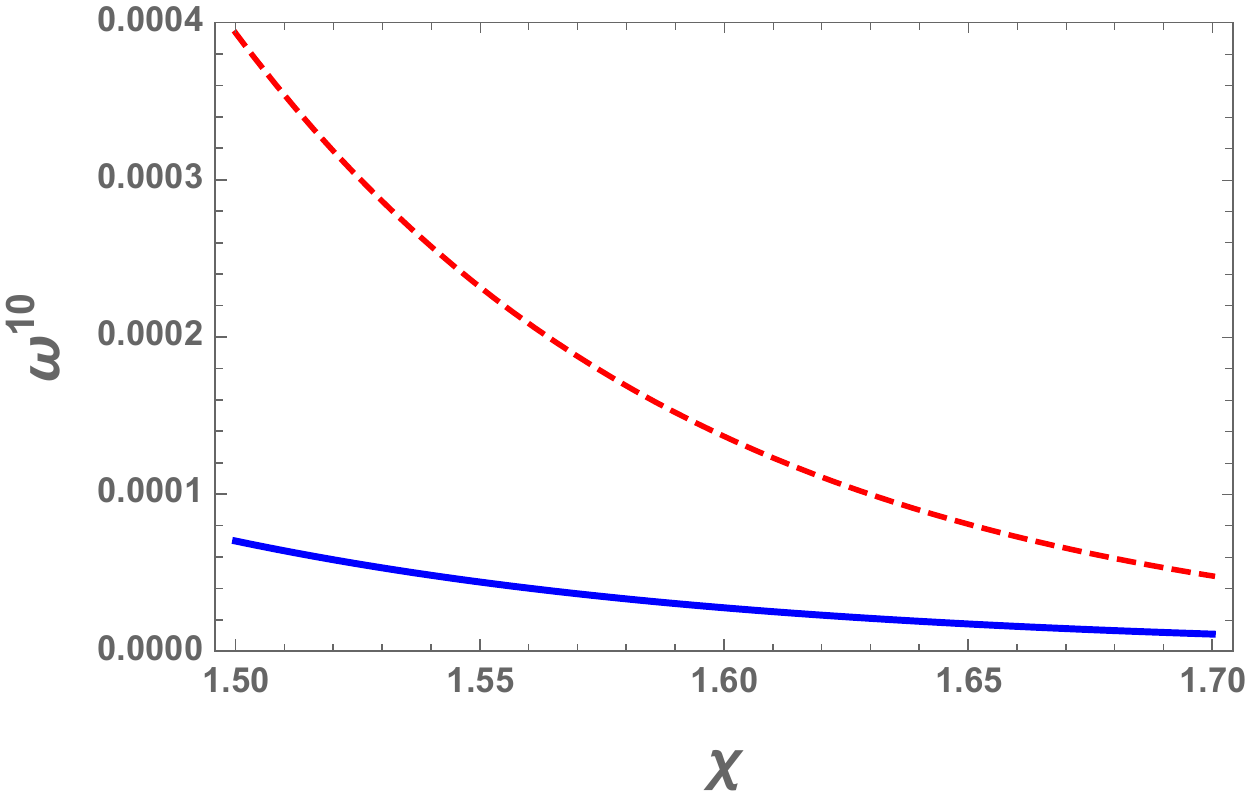}
\\
\end{tabular}
\end{center}
\vskip-.4cm
\caption{
The plot on the left shows the contribution arising from fermion zero modes $\omega_{\rm\,  ferm}$ for a single light flavour
(solid line). The dashed line is the large separation approximation $\frac{\sqrt{2}}{\left(1+\chi^2/2\right)^{3/2}}$.
The plot on the right shows
the corresponding contributions to the fermion prefactor ${\cal K}_{\rm ferm} $ in \eqref{eq:Kferm} for $N_f=5$.}
\label{fig:KF}
\end{figure}

In addition to the gauge-field interactions in the final state that are already accounted for by the semiclassical exponent 
$e^{- S_{I\bar{I}}}$ in  \eqref{eq:op_th}, there are also fermionic contributions to the final state. These arise from the $2N_f$
fermion zero modes in the amplitude in \eqref{eq:Gn} and give rise to the factor ${\cal K}_{\rm ferm}$ 
on the r.h.s. of \eqref{eq:op_th},
\begin{equation}
{\cal K}_{\rm ferm} \,=\, (\omega_{\rm\,  ferm})^{2N_f}\,, 
\label{eq:Kferm}
\end{equation}
where $\omega_{\rm\,  ferm}$ was computed at large separations in \cite{Balitsky:1992vs},
$\omega_{\rm\,  ferm} \simeq \frac{\sqrt{2}}{\left(1+\chi^2/2\right)^{3/2}}$, while the more general formula was derived 
in \cite{Ringwald:1998ek},
\begin{equation} 
\omega_{\rm\,  ferm} \,= \,  
\frac{3\pi}{8}\frac{1}{z^{3/2}}\,{}_2F_{1} \left(\frac{3}{2},\frac{3}{2};4;1-\frac{1}{z^2}\right)
\,,
\label{eq:omegaF}
\end{equation}
and this will be the expression that we will use. We plot $\omega_{\rm\,  ferm}(\chi)$ along with its large-$\chi$ approximation
in Fig.~\ref{fig:KF}. On the right plot we show the entire fermion prefactor ${\cal K}_{\rm ferm} $ for $N_f=5$.

The final ingredient appearing on the r.h.s. of \eqref{eq:op_th} is the product of four LSZ-reduced (anti-)instanton fields 
$A^{\rm inst}_{LSZ}(\pm p_i)$ for the 
two initial gluons with momenta $p_1$, $p_2$.
Starting from the instanton and anti-instanton solutions in the  coordinate space and Fourier-transforming it, we get after taking the on-shell  limit $p^2 \to 0$, 
\begin{eqnarray}
A_\mu^{a\, {\rm inst}}(x) &=& \frac{2 \rho^2}{g} \frac{\bar{\eta}^a_{\mu\nu} (x-x_0)_\nu}{(x-x_0)^2((x-x_0)^2+\rho^2)}
\, \longrightarrow \, A_\mu^{a\, {\rm inst}}(p)=\frac{4i\pi^2 \rho^2}{g} \frac{\bar{\eta}^a_{\mu\nu} p_\nu}{p^2} \, e^{ip \cdot x_0}\,\,\,\,\,
\label{eq:instFT}\\
A_\mu^{a\, {\overline{\rm inst}}}(x) &=& \frac{2 \bar\rho^2}{g} \frac{\bar{\eta}^a_{\mu\nu} (x-\bar{x}_0)_\nu}{(x-\bar{x}_0)^2((x-\bar{x}_0)^2+\bar\rho^2)}
\, \longrightarrow \, A_\mu^{a\, {\overline{\rm inst}}}(p) = \frac{4i\pi^2 \bar\rho^2}{g} \frac{{\eta}^a_{\mu\nu} p_\nu}{p^2} \, e^{ip \cdot \bar{x}_0}    
\end{eqnarray}   
Here $x_{0}$ and $\bar{x}_0$ are the instanton and anti-instanton centres, and $\bar{\eta}^a_{\mu\nu}$, ${\eta}^a_{\mu\nu}$, are the 't Hooft eta symbols \cite{tHooft:1976snw}. 
The LSZ reduction of the instanton configuration on the r. h. s of \eqref{eq:instFT} gives,
\begin{equation}
 A^{a\, {\rm inst}}_{LSZ}(p,\lambda)\,=\, \lim_{p^2\to 0} p^2 \epsilon^\mu(\lambda) \, A_\mu^{a\, {\rm inst}}(p)
 \,=\, \epsilon^\mu(\lambda) \,\bar{\eta}^a_{\mu\nu} p_\nu \, \frac{4i\pi^2 \rho^2}{g} \, e^{ip \cdot x_0}
 \,, \label{eq:Inst_LSZA}
\end{equation}
where $\epsilon^\mu(\lambda)$ is the polarisation vector for a gluon with a helicity $\lambda$. Using the identity,
\[
\sum_{\lambda=1,2}  \epsilon_\mu(\lambda)   \epsilon^*_\nu(\lambda) \,=\, - g_{\mu \nu}\,,
\]
and the properties of the 't Hooft eta symbols, we find for the pair of the gluon legs with the same incoming/outgoing momentum the expression,
 \begin{equation}
 \frac{1}{3}\sum_{a=1}^3 \, \frac{1}{2} \sum_{\lambda=1,2} \, A^{a\, {\rm inst}}_{LSZ}(p,\lambda)\, A_{LSZ}^{a\, {\overline{\rm inst}}}(-p;\lambda) \, =\,
 \frac{1}{6}\, 
 \left(\frac{2\pi^2}{g} \rho \bar\rho \,\sqrt{s'}\right)^2\, e^{iR\cdot p} \,,
\label{eq:LSZA1}
\end{equation}   
where $R=x_0-\bar{x}_0$ is the separation between the instanton--anti-instanton centres, and the factors 1/3 and 1/2 arise from averaging over the
three\footnote{The instanton and anti-instanton configurations we are suing live in the same SU(2) subgroup of the colour SU(3), hence we are summing the
't Hooft eta symbols over $a=1,2,3$ rather than $a=1,\ldots, 8$.}
SU(2) isospin components and two polarisations $\lambda$.

This reasoning leads to the following expression for the four external gluons appearing on the  r.h.s. of \eqref{eq:instFT}, 
\begin{equation}
 A^{\rm inst}_{LSZ}(p_1)\,A^{\rm inst}_{LSZ}(p_2)\,A^{\overline{\rm inst}}_{LSZ}(-p_1)\,A^{\overline{\rm inst}}_{LSZ}(-p_2) \,=\,
 \frac{1}{36}\, \left(\frac{2\pi^2}{g} \rho \bar\rho\, \sqrt{s'}\right)^4\, e^{iR\cdot (p_1+p_2)} \,.
\label{eq:LSZA}
\end{equation}    
The contribution $e^{iR\cdot (p_1+p_2)}$ arises from the
exponential factors $e^{ip_i\cdot x_0}$ and $e^{-ip_i\cdot \bar{x}_0}$ from the two instanton and two anti-instanton legs, which upon the Wick rotation to the 
Minkowski space becomes $e^{R_0 \sqrt{s'}}$. This concludes our overview of 
 of the ingredients appearing on the r.h.s. of \eqref{eq:op_th}.

Combining all these contributions allows us to express \eqref{eq:op_th} in the form,
\begin{eqnarray}
\hat\sigma_{\rm tot}^{\rm (cl)\,inst} &\simeq& \frac{1}{s'}\,{\rm Im}\, \frac{\kappa^2 \pi^4}{36\cdot4}   \int \frac{d\rho}{\rho^5}
 \int \frac{d\bar{\rho}}{\bar{\rho}^5}  \int d^4 R \int d\Omega  \,\left( \frac{2\pi}{\alpha_s(\mu_r)}\right)^{14} (\rho^2 \sqrt{s'})^2 (\bar{\rho}^2\sqrt{s'})^2\, {\cal K}_{\rm ferm}  \nonumber  \\
&& \,
(\rho \mu_r)^{b_0} (\bar{\rho} \mu_r)^{b_0}\, \exp\left(R_0 \sqrt{s'}\,-\,\frac{4\pi}{\alpha_s(\mu_r)}\,\hat{\cal S}(z)\right),
\label{eq:op_th2}
\end{eqnarray}

Note that \eqref{eq:op_th2} holds for general $\rho$ and $\bar\rho$ collective coordinates (no assumption is made about $\rho=\bar{\rho}$), they are independent integration variables. The factors ${\cal K}_{\rm ferm}(z) $ was defined in \eqref{eq:Kferm}-\eqref{eq:omegaF}
and $\hat{\cal S}(z)$ in \eqref{eq:Szdef}, both in terms of the conformal ratio $z$ that depends on $R,\rho, \bar\rho$ via \eqref{eq:zdef}.

We note that the expression on the r.h.s of \eqref{eq:op_th2} is of correct dimensionality ensured by the factor of $1/s$, 
with the remaining integral 
being dimensionless. The integrations over the collective coordinate $R_\mu$, $\rho$, $\bar{\rho}$ and $\Omega$ of the instanton--anti-instanton 
configuration are to be carried in the steepest descent approach, i.e.\ by finding the saddle-point extremum of the expression in the
 exponent. 
It is easy to  see 
 that the relative $I \bar{I}$ separation $R=|R_\mu|$ collective coordinate gives rise to a single negative mode of the quadratic fluctuation operator expanded around the saddle point in the exponent of \eqref{eq:op_th2}. Indeed, for fixed values of (anti)-instanton sizes, there is a competition between the positive factor
 ${R_0 \sqrt{s'}}$ that grows with $R_0$ and the negative-valued factor $- \hat{\cal S}(z)$ which leads to the exponential suppression at 
 large $R_0$.\footnote{ The dependence of the $I \bar{I}$ action on $R/\rho$ is shown in Fig.~\ref{fig:S}.}
This results in the saddle point of the exponent along the $R_0$ direction with $R_{1,2,3}=0$.  Carrying out the Gaussian integrations over 
 the fluctuations around the saddle-point (the task we perform in the following subsection) will result in an imaginary-valued expression, thus furnishing the required 
 imaginary part of the integral in \eqref{eq:op_th2} as required by the optical theorem  \cite{Khoze:1991mx}. 
 We will confirm that this is indeed the case by evaluating the determinant of the relevant second derivatives operator in Eq.~\eqref{eq:detK}.

It is well-known, however, that the expression for the cross-section in \eqref{eq:op_th2} suffers from a severe infrared problem
arising from instantons of large size, $\rho \to \infty$. In QCD, unlike the electroweak theory, there are no  scalar fields whose VEVs 
would cut off integrations over large $\rho$ in \eqref{eq:op_th2}. The expression in \eqref{eq:op_th2} was obtained using the
 leading-order semiclassical expansion around the instanton--anti-instanton configuration. 
 At the classical level, QCD is of course scale-invariant, so there is no surprise that the leading-order semiclassical 
 expression does not fix the instanton size.
 To break classical scale-invariance we need to 
 include quantum corrections that describe interactions of the initial state gluons. This corresponds to allowing for 
 fluctuations around the four (anti)-instanton fields appearing in front of the exponent in \eqref{eq:op_th}. This   
 amounts to inserting propagators in the instanton background between pairs  of gluon fields in the pre-exponential factor 
 in \eqref{eq:op_th}   and re-summing the resulting perturbation theory. This programme has been carried out by Mueller 
 in \cite{Mueller:1990qa,Mueller:1990ed}. It was shown that the quantum corrections due to interactions of the initial 
 states exponentiate and the resulting expression for 
 the resummed quantum corrections 
  gives the factor 
 $e^{- \alpha_s \, \rho^2 s' \log s'}$ for the instanton, and the analogous factor for the anti-instanton
 in the optical theorem expressions \eqref{eq:op_th} and \eqref{eq:op_th2}. 
  
 We thus obtain the quantum-corrected expression for the instanton production cross-section,
 \begin{eqnarray}
\hat\sigma_{\rm tot}^{\rm inst} &\simeq& 
\frac{1}{s'}\,{\rm Im}\, \frac{\kappa^2 \pi^4}{36\cdot4}   \int \frac{d\rho}{\rho^5}
 \int \frac{d\bar{\rho}}{\bar{\rho}^5}  \int d^4 R \int d\Omega  \left( \frac{2\pi}{\alpha_s(\mu_r)}\right)^{14} (\rho^2 \sqrt{s'})^2 (\bar{\rho}^2\sqrt{s'})^2\, {\cal K}_{\rm ferm}
\nonumber
  \\
&& \,
 (\rho \mu_r)^{b_0} (\bar{\rho} \mu_r)^{b_0}\,\exp\left(R_0 \sqrt{s'}\,-\,\frac{4\pi}{\alpha_s(\mu_r)} \,\hat{\cal S}(z)
\,-\, \frac{\alpha_s(\mu_r)}{16\pi}(\rho^2+\bar{\rho}^2) \, s'\, \log \left(\frac{s'}{\mu_r^2}\right)
\right).\nonumber\\
 \label{eq:op_th3}
\end{eqnarray}

The expression \eqref{eq:op_th3} is the key technical input on which the results this paper are based.
It combines the semi-classical instanton contribution to the total cross-section including 
the effects of final state interactions derived in Ref.~\cite{Khoze:1991mx},
with the resummed quantum corrections in the initial state that were computed by Mueller in Ref.~\cite{Mueller:1990ed}.
It is easily verified that the initial state interactions quantum effect provides an exponential  cut-off 
of the large instanton/anti-instanton sizes; the cut-off scale is set by the (partonic) energy scale $s'\log s'$ of the scattering
process, and further it contains a factor of $\alpha_s$, as it should in the radiative corrections.

\subsection{The saddle-point solution and the instanton cross-section}
\label{sec:saddle}
 
Now we can search for the saddle-point in $R_\mu$, $\rho$ and $\bar{\rho}$ that extremises the function in the exponent 
in \eqref{eq:op_th3}. The instanton--anti-instanton separation coordinate is stabilised along the $R_0$ direction due to the
interplay between the $R_0 \sqrt{s'}$ and $-\,\frac{4\pi}{\alpha_s(\mu_r)} \,\hat{\cal S}(z)$ factors in the exponent. 
The saddle-point is at $R=R_0$, and to simplify our notation we will re-write the first term as $R \sqrt{s'}$ at the saddle-point.
Furthermore, the symmetry between the instanton and anti-instanton configuration in the forward elastic scattering amplitude implies
that the saddle-point value of $\rho$ will be equal to $\bar{\rho}$.\footnote{We checked numerically that there is a saddle-point solution 
with $\rho=\bar{\rho}$. This does not exclude the logical possibility that there may exist additional pairs of saddle-points on which the $Z_2$ symmetry between the instanton and the anti-instanton is broken spontaneously, i.e. $\{ \rho=A,\bar\rho=B\}$ and $\{ \rho=B,\bar\rho=A\}$. We have not investigated this in detail. If such new saddle-points are present, they may provide additional semiclassical contributions }

 So, in obtaining the saddle-point solution, we can set
$\bar{\rho}=\rho$ and search for the extremum of the `holy-grail' function,
\begin{equation}
{\cal F}\,=\,
R \sqrt{s'}\,-\,\frac{4\pi}{\alpha_s(\mu_r)} {\cal S}(R/\rho)
\,-\, \frac{\alpha_s(\mu_r)}{8\pi}\rho^2s'\, \log (s'/\mu_r^2)\,,
\label{eq:fhg}
\end{equation}
that appears in the exponent in \eqref{eq:op_th3}.

To emphasise the applicability of the saddle-point approximation to the integral \eqref{eq:op_th3}, we chose the rescaled dimensionless 
integration variables,
\begin{equation}
\tilde\rho\,=\, \frac{\alpha_s(\mu_r)}{4\pi} \, \sqrt{s'} \rho\,,
\qquad
\chi\,=\, \frac{R}{\rho}\,,
\label{eq:resc}
\end{equation}
and write the holy-grail function \eqref{eq:fhg} as,
\begin{equation}
{\cal F}\,=\, \frac{4\pi}{\alpha_s(\mu_r)} \, F(\tilde\rho, \chi)
\,, \qquad
F\,=\,
\tilde\rho \,\chi\,-\, {\cal S}(\chi)
\,-\,  \tilde\rho^2 \, \log (\sqrt{s'}/\mu_r)\,.
\label{eq:fhg2}
\end{equation}
Instanton calculations are based on a semi-classical approach that is valid in a weak-coupling regime, hence the overall
factor 
$\frac{4\pi}{\alpha_s(\mu_r)} \gg 1$ in front of $F$ justifies the steepest descent approach where the integrand in \eqref{eq:op_th3}
is dominated by the 
the saddle-point of $F(\tilde\rho, \chi)$ in \eqref{eq:fhg2}.

Before proceeding to solve the saddle-point equations that extremise the holy-grail function $F$ above, 
we would like to comment on how to select the value of the renormalisation scale $\mu_r$.
Recall that the integrand in \eqref{eq:op_th3} contains the factor,
\begin{equation}
(\rho \mu_r)^{b_0} (\bar{\rho} \mu_r)^{b_0}\, 
e^{-\,\frac{4\pi}{\alpha_s(\mu_r)}} \,=\, e^{-\,\frac{2\pi}{\alpha_s(1/\rho)}-\,\frac{2\pi}{\alpha_s(1/\bar{\rho})}}\,,
\label{eq:RGinv}
\end{equation}
where $(\rho \mu_r)^{b_0}$ and $ (\bar{\rho} \mu_r)^{b_0}$ come from the instanton and the anti-instanton measure $D(\rho)$ and  $D(\bar{\rho})$, and
the factor $e^{-\,\frac{4\pi}{\alpha_s(\mu_r)}}$ accounts for the instanton and the anti-instanton action contributions in the dilute limit.
The r.h.s. of \eqref{eq:RGinv} is RG-invariant at one-loop, it does not depend on the choice of $\mu_r$, instead the scale of the running coupling constant is set 
at the inverse instanton and anti-instanton sizes. 

There are two methods for fixing the RG scale that one can follow; they both should give equivalent results at the level of accuracy our semi-classical 
instanton approach provides. 
\begin{enumerate}
\item The first method is to solve the saddle-point equations keeping $\mu_r $ fixed. The saddle-point equations 
 $\partial_\chi F=0$ and $\partial_{\tilde{\rho}} F=0$ arise from extremising the function
\begin{equation}
F\,=\,
\tilde\rho \,\chi\,-\, {\cal S}(\chi)
\,-\,  \tilde\rho^2 \, \log (\sqrt{s'}/\mu_r)
\,+\, 2b_0  \frac{\alpha_s(\mu_r)}{4\pi} \log(\rho\mu_r)\,.
\label{eq:fhg2mu}
\end{equation}
Then after finding the saddle-point solution for $\chi$ and $\tilde{\rho}$ we set $\mu_r = 1/\rho$ at the saddle-point value. 
Note that we have added the last term on the r.h.s. of \eqref{eq:fhg2mu} to account for the back reaction of the $(\rho \mu_r)^{b_0} (\bar{\rho} \mu_r)^{b_0}$
factor on the saddle-point. Of course, after setting $\mu_r = 1/\rho$  in the $F$ computed at the saddle-point, this term disappears.

\item The alternative approach is set $\mu_r=1/\rho$ from the beginning. 
The function in the exponent is \eqref{eq:op_th3} (note that we do not pull out the $4\pi/\alpha_s(\rho)$ factor),
\begin{equation}
{\cal F}\,=\, 
\rho\chi \sqrt{s'} \,-\, \frac{4\pi}{\alpha_s(\rho)}\, {\cal S}(\chi) \,-\, \frac{\alpha_s(\rho)}{4\pi}\,\rho^2 s' \, \log (\sqrt{s'}\rho)\,.
\label{eq:Ffullrho}
\end{equation}
We look for the saddle-point solutions of the equations $\partial_\chi {\cal F}=0$ and $\partial_{\rho} \,{\cal F}=0$ for the variables $\chi$ and $\rho$.
\end{enumerate}

\noindent We have computed the instanton production cross-sections following both of these methods and have found that the numerical results 
for $\hat\sigma_{\rm tot}^{\rm inst}$ as the function of $\sqrt{s'}$ are 
in good agreement with each other. This demonstrates that our approach is stable against such variations in the RG scale selection procedure.

In what follows we will concentrate on the second method where all the couplings are from the beginning taken at the scale 
set by the characteristic instanton size.
We now solve the saddle-point equations $\partial_\chi {\cal F}=0$ and $\partial_{\rho} \,{\cal F}=0$ for \eqref{eq:Ffullrho} and find,
\begin{equation}
\rho\sqrt{s'} \,=\, \frac{4\pi}{\alpha_s(\rho)}\, \frac{d{\cal S}(\chi)}{d\chi}\,,
\label{eq:sp1}
\end{equation}
and
\begin{equation}
\chi\,=
\, \frac{\alpha_s(\rho)}{4\pi}\, \rho\sqrt{s'}\, \left(2\log(\rho\sqrt{s'})+1\right)
\,+\, 2b_0 \left(\frac{\alpha_s(\rho)}{4\pi}\right)^2 \rho\sqrt{s'}\, \log(\rho\sqrt{s'})
\, -\, \frac{2b_0}{\rho\sqrt{s'}} {\cal S}(\chi),
\label{eq:sp2}
\end{equation}
where we made use of the one-loop RG relation for the derivative of the running coupling,
\begin{equation}
\partial_\rho\left(\frac{4\pi}{\alpha_s(\rho)}\right)\,=\, -\frac{2 b_0}{\rho}\,, \qquad
\partial_\rho\left(\frac{\alpha_s(\rho)}{4\pi}\right)= \left(\frac{\alpha_s(\rho)}{4\pi}\right)^2 \frac{2 b_0}{\rho}
\,.
\end{equation}
Our procedure for solving the saddle-point equations \eqref{eq:sp1}-\eqref{eq:sp2} is as follows. 
We introduce the already familiar rescaled variable $\tilde{\rho} = \frac{\alpha_s(\rho)}{4\pi} \, \sqrt{s'} \rho$, along with the new 
scaling parameter, 
\begin{equation}
u\,=\, \sqrt{s'} \rho\,,
\label{eq:udef}
\end{equation}
and write \eqref{eq:sp1}-\eqref{eq:sp2} as,
\begin{eqnarray}
\tilde{\rho} &=& {\cal S}'(\chi)\,, 
\label{eq:sp1t} \\
\chi&=& \tilde{\rho} \left(2 \log u +1\right)
\, +\, 2b_0\,\tilde{\rho}^2\, \frac{\log u}{u} 
\, -\, \frac{2b_0}{u} \,{\cal S}(\chi).
\label{eq:sp2t}
\end{eqnarray}
There are two saddle-point equations \eqref{eq:sp1}-\eqref{eq:sp2} to solve, to determine the two variables
$\tilde\rho$ and $\chi$ in \eqref{eq:resc}.
Their values as well as the final result for the
instanton cross-section of course depend on the energy $\sqrt{s'}$, which plays the role of the external input parameter.
In practice, instead of $\sqrt{s'}$ it is more convenient to characterise the process by the dimensionless input variable $u$ defined in \eqref{eq:udef}.

 \begin{figure}[]
\begin{center}
\begin{tabular}{cc}
\hspace{-.4cm}
\includegraphics[width=0.5\textwidth]{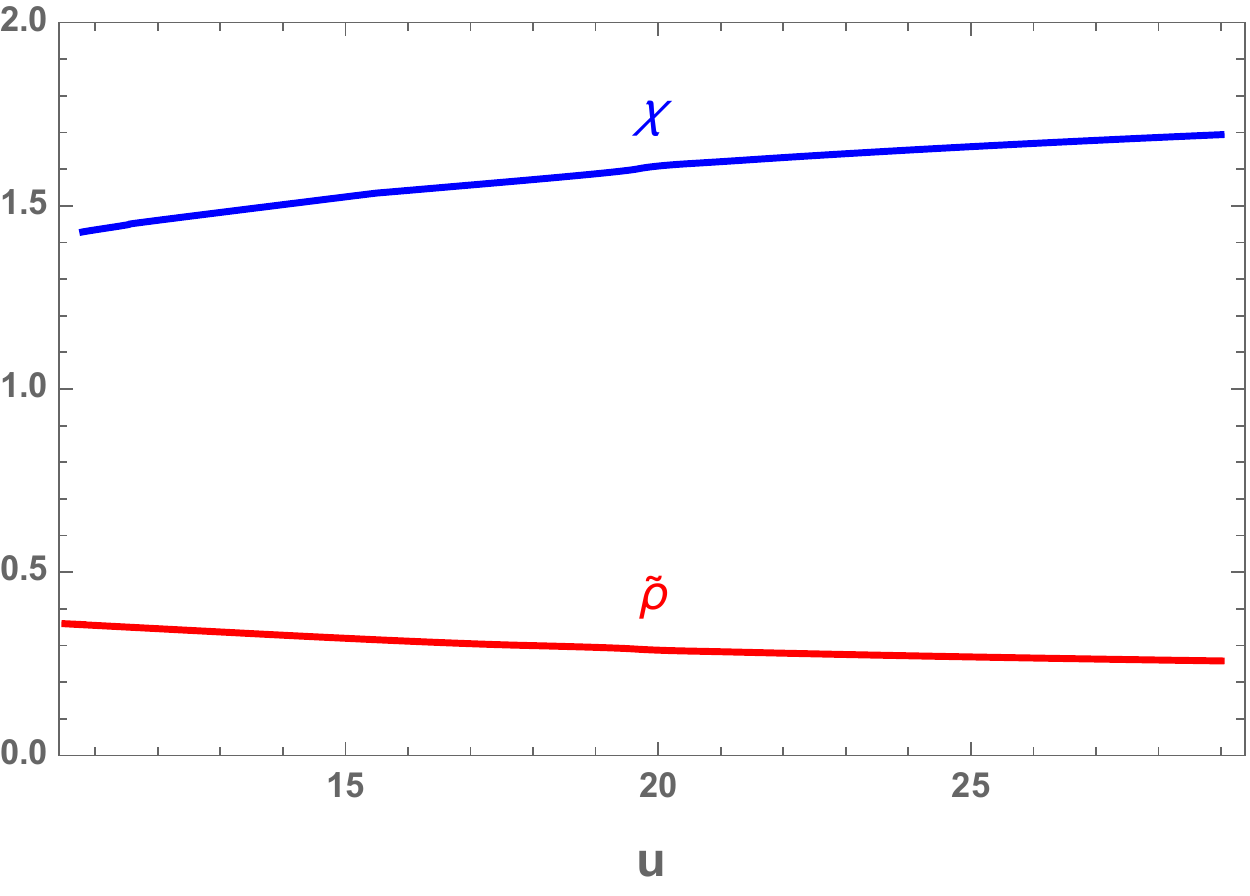}
&
\includegraphics[width=0.5\textwidth]{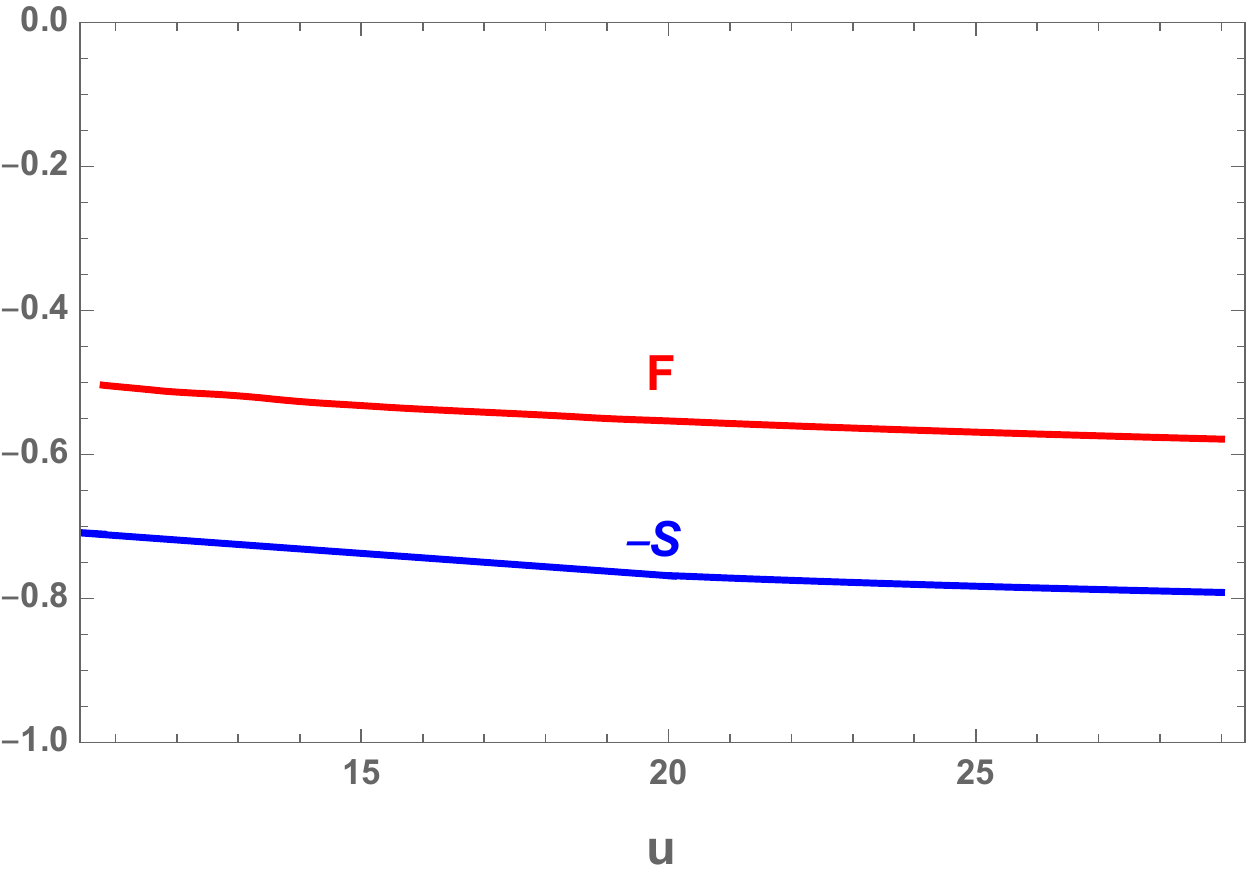}
\\
\end{tabular}
\end{center}
\vskip-.4cm
\caption{
The plot on the left shows the saddle-point solutions $\chi$ and $\tilde{\rho}$ as functions of the input variable $u$.  
The plot on the right gives the values of the holy-grail function $F= \frac{\alpha_s}{4\pi}{\cal F}$ and of the  (minus) instanton action $-{\cal S}(\chi)$ 
as functions of $u$.}
\label{fig:saddle}
\end{figure}

In summary, for every value of $u$ we solve the equations \eqref{eq:sp1}-\eqref{eq:sp2} numerically to find the saddle-point values of $\tilde{\rho}$ and $\chi$.
These are shown in Fig.~\ref{fig:saddle} along with the values of the holy grail function ${\cal F}$,
\begin{equation}
{\cal F}\,=\, \frac{4\pi}{\alpha_s(\rho)} \, \left(\tilde\rho \,\chi\,-\, {\cal S}(\chi)
\,-\,  \tilde\rho^2 \, \log u \right)\,.
\label{eq:fhg_fin}
\end{equation}
and the
 instanton-anti-instanton action ${\cal S}(\chi)$.
The corresponding (unrescaled) instanton size $\rho$ and the running coupling $\alpha_s(\rho)$  are obtained via,
\begin{equation}
\frac{4\pi}{\alpha_s(\rho)}\,=\, \frac{u}{\tilde{\rho}}\,, \qquad
\rho^{-1} \,=\, M_Z  \,e^{\frac{1}{b_0} \left(\frac{2\pi}{\alpha_s(\rho)} - \frac{2\pi}{\alpha_s(M_Z)}\right)} \,.
\label{eq:urho}
\end{equation}
From this we recover the $\sqrt{s'}$,
\begin{equation}
\sqrt{s'} \,=\, \rho^{-1} u\,.
\end{equation}

\begin{figure}[]
\begin{center}
\begin{tabular}{cc}
\hspace{-.4cm}
\includegraphics[width=0.5\textwidth]{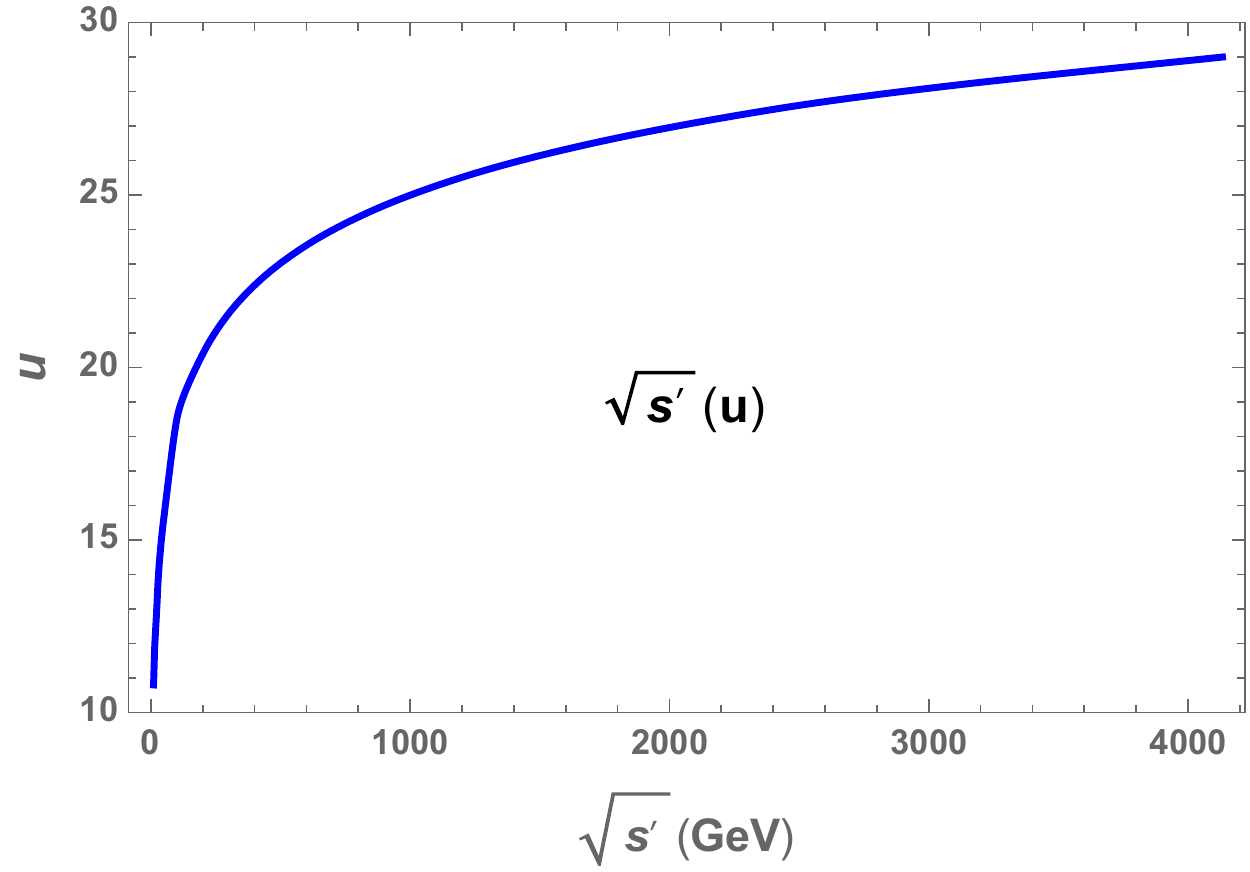}
&
\includegraphics[width=0.5\textwidth]{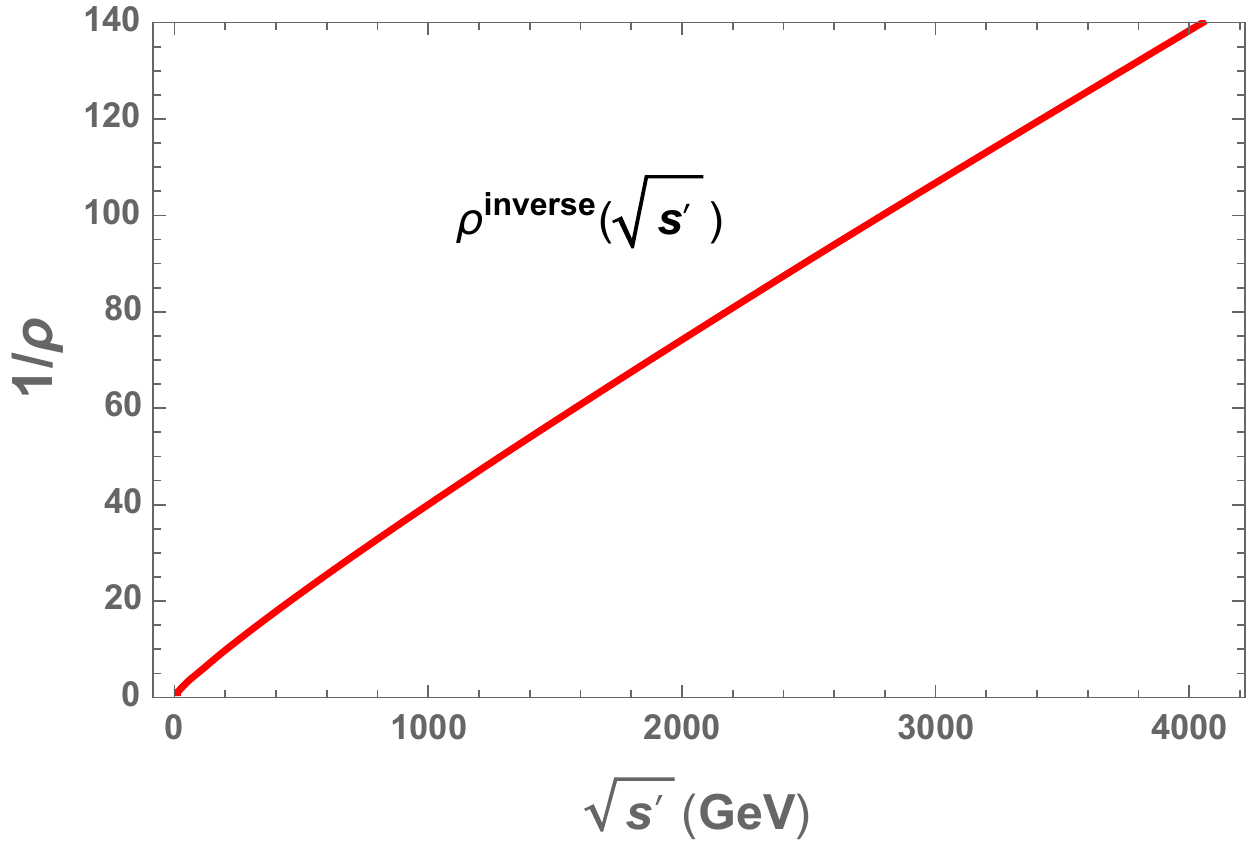}
\\
\end{tabular}
\end{center}
\vskip-.4cm
\caption{
The plot on the left shows $\sqrt{s'}$ measured in GeV and the function of the input variable $u$.  
The plot on the right gives the inverse instanton size (in GeV) as the function of energy}
\label{fig:surh}
\end{figure}

\begin{figure}[]
\begin{center}
\begin{tabular}{cc}
\hspace{-.4cm}
\includegraphics[width=0.5\textwidth]{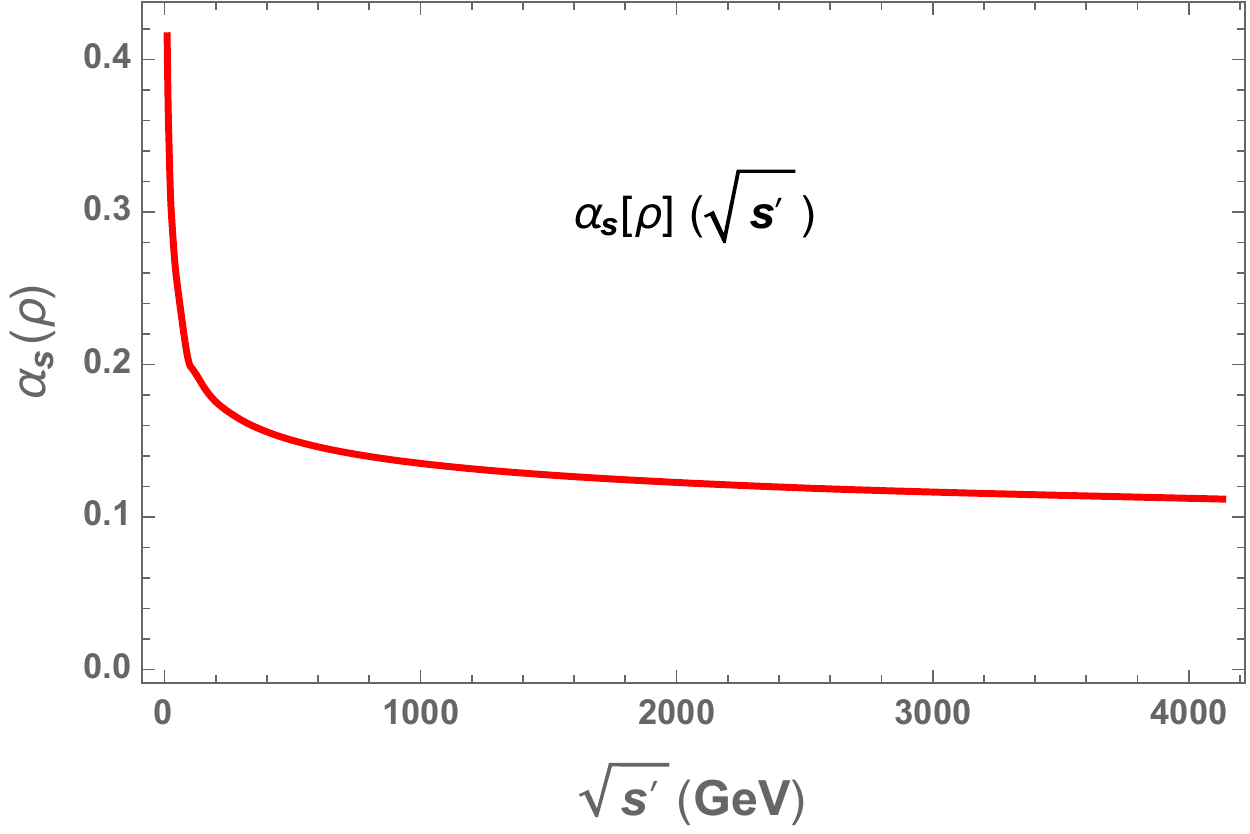}
&
\includegraphics[width=0.5\textwidth]{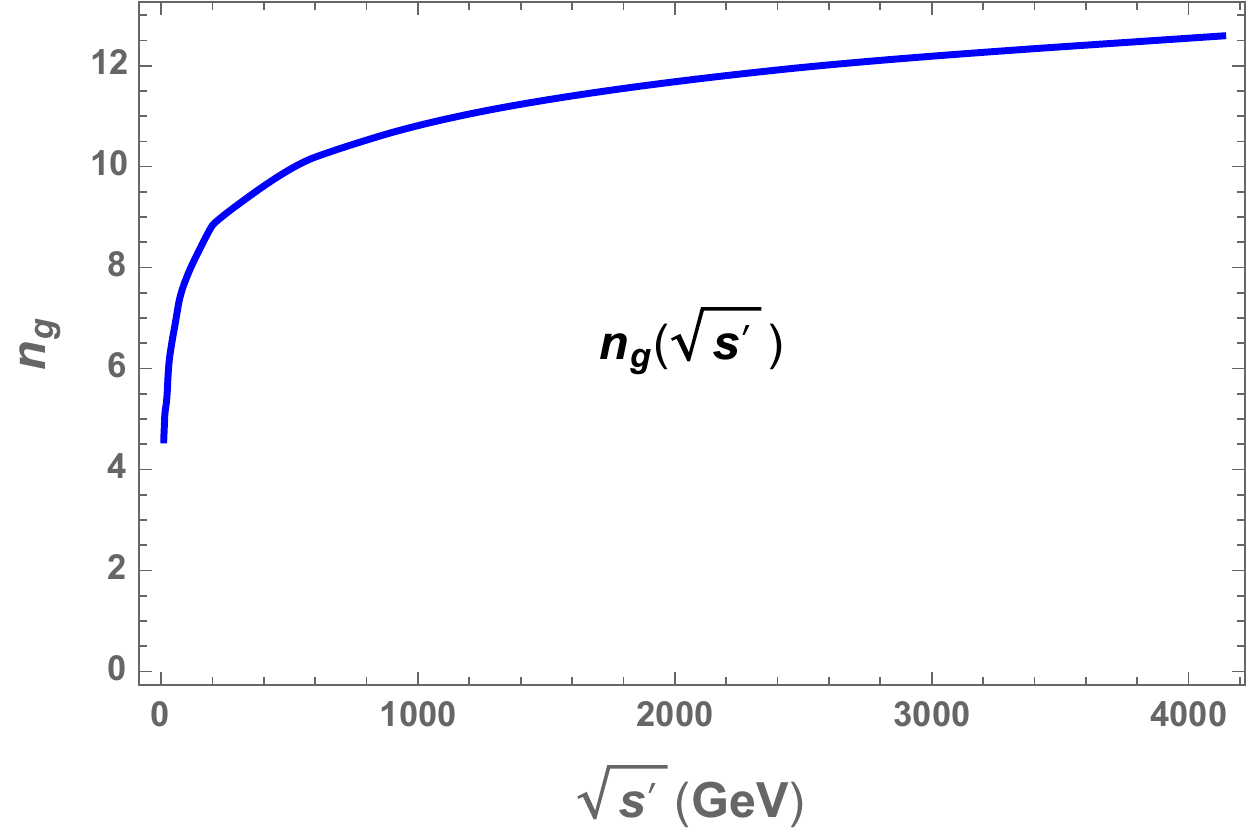}
\\
\end{tabular}
\end{center}
\vskip-.4cm
\caption{
The coupling constant $\alpha_s(\rho)$ for the instanton process of the instantons size $\rho$ as the function of energy (left plot). The plot on the right
shows the mean number of gluons, $n_g$,  in the final state.
}
\label{fig:alpng}
\end{figure}

We illustrate this procedure in Figs.~\ref{fig:surh} and \ref{fig:alpng}. The plot on the left in Fig.~\ref{fig:surh} shows the correspondence between the
input variable $u$ and the energy $\sqrt{s'}$ in GeV. The plot on the right shows the characteristic values of the inverse instanton size $1/\rho$ in GeV 
as the function of $\sqrt{s'}$.
The dependence of the coupling constant $\alpha_s (\rho)$ on the energy scale $\sqrt{s'}$ is plotted on the left graph of Fig.~\ref{fig:alpng}.
The right hand side plot of that figure shows the mean number of gluons $n_g$ in the final state, computed using Eq.~\eqref{eq:ng_n} below.

The mean number of gluons produced in the final state of the instanton process is easy to determine from the amplitude for the leading order-instanton process
\begin{equation} 
n_g\,=\, \frac{2\pi}{\alpha_s(\rho)} \,\, \tilde{\rho}^2\,  \log(u) \,.
\label{eq:ng_n}
\end{equation}
Indeed concentrating on the $n_g$ dependence of the integral over the instanton size for the $2\to n_g$ amplitude, we have,
\begin{eqnarray} 
{\cal A}_{\, 2\to\, n_g} &\sim& \int d\rho\,  \left(A_{\mu\,\,{\rm LSZ}}^{\rm inst}\right)^{n_g} \, e^{-S_I - \frac{\alpha_s(\rho)}{16\pi} s\rho^2 \log (s\rho^2)}
\nonumber \\
&\sim& e^{-S_I}\,  \int d\rho\,\, (\rho^2)^{n_g}\, e^{ - \frac{\alpha_s(\rho)}{16\pi} s\rho^2 \log (s\rho^2)}
 \,. 
\label{eq:ng_nn}
\end{eqnarray}
Next, by differentiating the integrand with respect to $\rho^2$ we identify the dominant contribution to the integral as coming from the solution of the extremum equation,
$ n_g/\rho^2= \alpha_s(\rho)/(16\pi)\, s \log (s\rho^2)$, which 
gives,
\begin{equation} 
n_g\,=\, \rho^2 s' \, \frac{\alpha_s(\rho)}{16 \pi} \,  \log(s'\rho^2) \,=\, \tilde \rho^2\, \frac{2\pi}{\alpha_s(\rho)}\, \log(u)\,=\,\,
\tilde \rho\, \frac{u\, \log(u)}{2}
\,.
\label{eq:ng_nfin}
\end{equation}
The second equality in the expression above reproduces Eq.~\eqref{eq:ng_n} we quoted above, and the last equality makes use of the first
equation in \eqref{eq:urho}.

The relation  \eqref{eq:ng_n}, \eqref{eq:ng_nfin} between the number of gluons and the dominant value of $\tilde{\rho}$  was obtained in the leading-order 
semiclassical approximation, but the saddle-point value of $\tilde \rho$ of course takes into account effects of the
final-state gluon interactions.
Numerical values for the mean number of gluons varies between between $n_g\simeq 5$ and $n_g\simeq 13$ when the energy $\sqrt{s'}$ varies over the 
broad range $10 {\rm GeV} < \sqrt{s'} < 4 {\rm TeV}$.

\medskip

The final task left to us before we can compute the instanton cross-section is to carry out the integrations in on the r.h.s. of \eqref{eq:op_th3}
around the saddle-point value for $\tilde{\rho}$ and $\chi$.
Integrations over the spatial components of the $I\bar{I}$ separation $\int d^3 {\bf R}$ contribute the term ${\cal P}_{\bf \, R}$
to the pre-factor in the cross-section, where
\begin{equation}
{\cal P}_{\bf\, R} \,=\, \left(\frac{\alpha_s(\rho) \, \chi }{2{\cal S}'(\chi) }\right)^{3/2} \rho\,.
\label{eq:PRdef}
\end{equation}
The integration over the difference between the instanton and anti-instanton sizes, $\delta=\rho-\bar{\rho}$,
gives (where in the second equality we used $\tilde{\rho}={\cal S}'(\chi)$),
\begin{equation}
{\cal P}_{\delta} \,=\, \left(\frac{\alpha_s(\rho)  }{\frac{4+\chi^2}{2\chi}{\cal S}'(\chi) \,+\, \tilde{\rho}^2 \log u}\right)^{1/2} \rho
\,=\, \left(\frac{\alpha_s(\rho) \,2\chi }{{\cal S}'(4+\chi^2\,+\, {\cal S}' \log u)}\right)^{1/2} \rho
\,.
\end{equation}
The integrations over the relative orientations $\Omega$ around the maximally attractive value at the saddle-point, are
can also be straightforwardly carried out following Ref.~\cite{Balitsky:1992vs},  with the result,
\begin{equation}
{\cal P}_{\,\Omega} \,=\, \frac{3\sqrt{3}}{\pi^4} \left(\frac{\alpha_s(\rho)  \,\chi}{(2+\chi^2)\, {\cal S}'(\chi)}\right)^{7/2}.
\label{eq:PUdef}
\end{equation}
Finally, the integral over the two remaining variables gives,
\begin{eqnarray}
{\cal P}_{\chi\tilde\rho}\,:=\,
{\rm Im}\, \int dR\,  d\rho\, e^{{\cal F}(\chi,\tilde{\rho})} &=&
\frac{4\pi}{\alpha_s(\rho)} \frac{\rho}{\sqrt{s'}}\,\,  {\rm Im}\, \int d\chi\,  d\tilde{\rho} \, e^{{\cal F}(\chi,\tilde{\rho})} \nonumber\\
&=& \frac{4\pi}{\alpha_s(\rho)} \frac{\rho}{\sqrt{s'}}\,\,  {\rm Im}\, \frac{2\pi}{{\rm det}^{1/2} K},
\label{eq:prefRrho}
\end{eqnarray}
where ${{\rm det}^{1/2} K}$ is the square root of the determinant of the matrix $K$ of second derivatives of $-{\cal F}(\chi,\tilde{\rho})$ 
with respect to $\chi$ and $\tilde\rho$,
\begin{eqnarray}
K\,:=\, -\, \frac{\partial^2 {\cal F}}{(\partial \tilde{\rho}, \partial \chi)}
\,=\, \frac{4\pi}{\alpha_s(\rho)}
\begin{pmatrix} 
2 \log u & -1 \\
-1 & {\cal S}^{\prime\prime}(\chi) 
\end{pmatrix}\,, 
\end{eqnarray}
so that,
\begin{eqnarray}
{\rm det}\,K\,=\, -\left(\frac{4\pi}{\alpha_s(\rho)}\right)^2
\left(1+ (-{\cal S}^{\prime\prime}(\chi) 2\log u \right)\,.
\label{eq:detK}
\end{eqnarray}
Note that the determinant is negative-valued (the quantity $-{\cal S}^{\prime\prime}(\chi) >0$), thus its square root does indeed contribute to the imaginary
part of the expression on the r.h.s. \eqref{eq:prefRrho}.
As the result, integral \eqref{eq:prefRrho} gives,
\begin{eqnarray}
{\cal P}_{\chi\tilde\rho}\,=\,
2\pi\, \frac{\rho}{\sqrt{s'}} \, \frac{1}{ \left(1+ (-{\cal S}^{\prime\prime}(\chi) 2\log u \right)^{1/2}}.
\label{eq:Pchirho}
\end{eqnarray}

\medskip

 Assembling all the contributions listed above in \eqref{eq:fhg_fin}, \eqref{eq:PRdef}-\eqref{eq:PUdef}, \eqref{eq:Pchirho}
we find for the total parton-level instanton cross-section \eqref{eq:op_th3},
the following expression,
\begin{equation}
\hat\sigma_{\rm tot}^{\rm inst} \,=\, \frac{1}{s'}\, {\cal P}\, e^{\,{\cal F}}\,,
\label{eq:sigmafin}
\end{equation}
where
\begin{eqnarray}
{\cal F}
&=& \frac{4\pi}{\alpha_s(\rho)} \, \left(\tilde\rho \,\chi\,-\, {\cal S}(\chi)
\,-\,  \tilde\rho^2 \, \log u \right)\,,
\label{eq:fhg_fin2}
 \\
{\cal P} &=&\,
\frac{\kappa^2 \, (2/3) \sqrt{6} \,\pi^{13/2} \, 
\left( \frac{2\pi}{\alpha_s(\rho)}\right)^{17/2} \chi^{11/2}\,u^3\,  {\cal K}_{\rm ferm}}{(2+\chi^2)^{7/2}({\cal S}')^{11/2}
\left(4+\chi^2+2\chi {\cal S}' \log u\right)^{1/2}
 \left(1+(-2{\cal S}^{\prime\prime})\log u \right)^{1/2}}.
 \label{eq:op_Pfinal}
\end{eqnarray}
Our result for the prefactor in \eqref{eq:op_Pfinal} can be further simplified and re-written as a function of just two variables, $\chi$ and $u$, 
with the help of \eqref{eq:sp2t} and \eqref{eq:urho},
\begin{equation}
\tilde\rho= {\cal S}'(\chi)\,, \qquad
\frac{4\pi}{\alpha_s(\rho)}\,=\, \frac{u}{2 {\cal S}'(\chi)}\,,
\label{eq:2urho}
\end{equation}
with the result,
\begin{eqnarray}
{\cal P} &=&\,
\frac{\kappa^2 \, \pi^{13/2} \, \chi^{11/2}\,u^{23/2}\, {\cal K}_{\rm ferm}}{2^7 \sqrt{3}\, ({\cal S}')^{14} (2+\chi^2)^{7/2}
\left(4+\chi^2+2\chi {\cal S}' \log u\right)^{1/2}
 \left(1+(-2{\cal S}^{\prime\prime})\log u \right)^{1/2}}. \,\,\,\,\,
 \label{eq:op_Pfinal2}
\end{eqnarray}

 \begin{figure}[]
\begin{center}
\begin{tabular}{cc}
\hspace{-.4cm}
\includegraphics[width=0.5\textwidth]{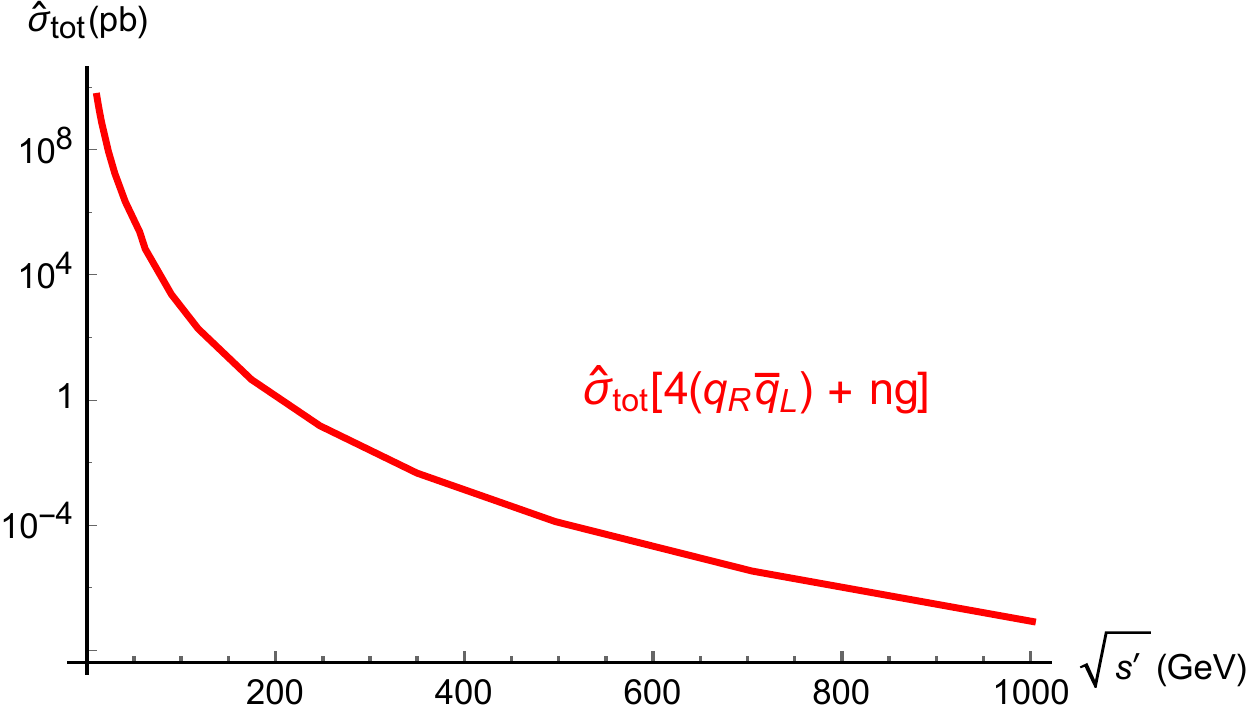}
&
\includegraphics[width=0.5\textwidth]{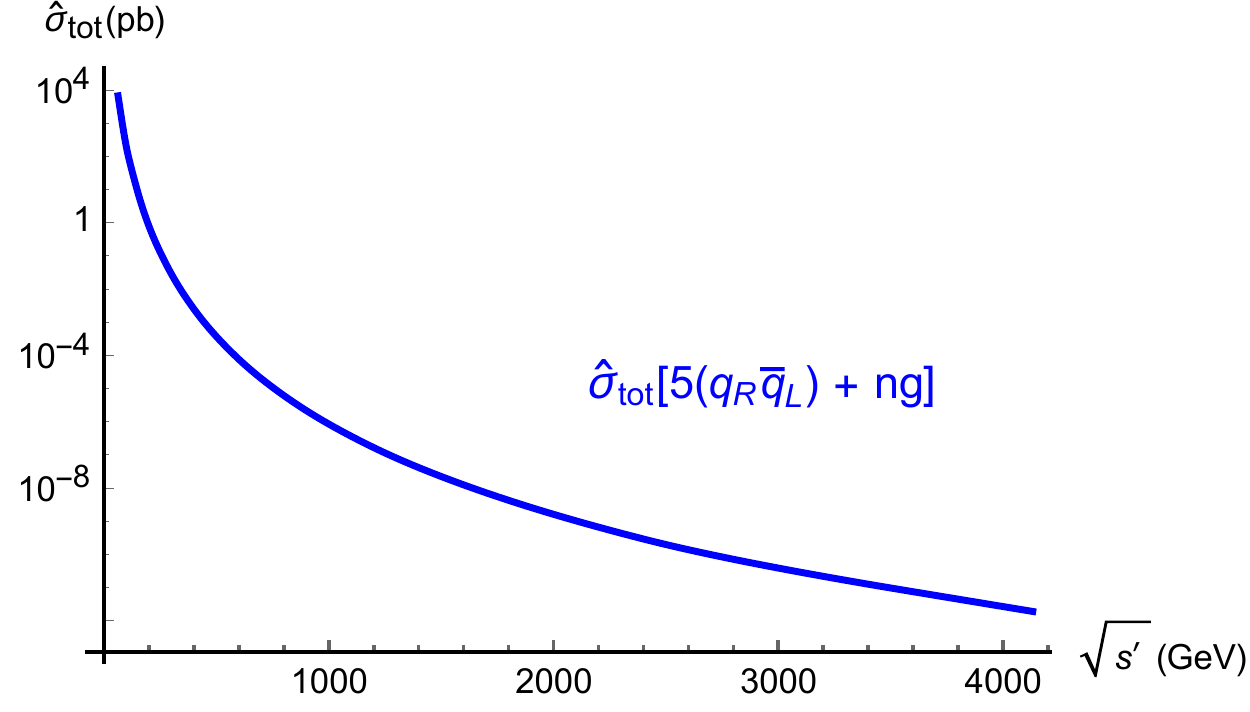}
\\
\end{tabular}
\end{center}
\vskip-.4cm
\caption{
Instanton cross-section $\hat\sigma_{\rm tot}^{\rm inst}$ as the function of partonic CoM energy $\sqrt{s'}$. The plot on the left is
for eight $q\bar{q}$ pairs in the final state, and the plot on the left is for ten $q\bar{q}$ pairs. The number of final state gluons is general,
with the mean given by Eq.~\eqref{eq:ng_n}.
}
\label{fig:sigma}
\end{figure}

 \begin{figure}[]
\begin{center}
\begin{tabular}{cc}
\hspace{-.4cm}
\includegraphics[width=0.5\textwidth]{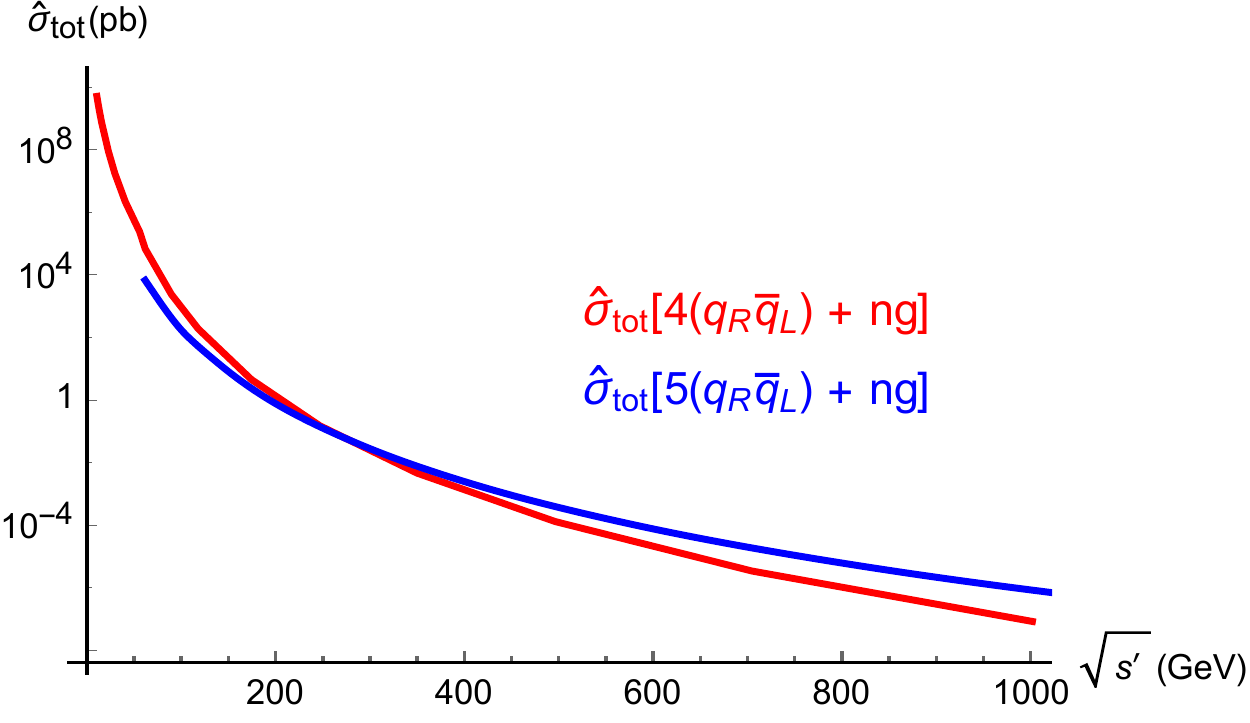}
&
\includegraphics[width=0.5\textwidth]{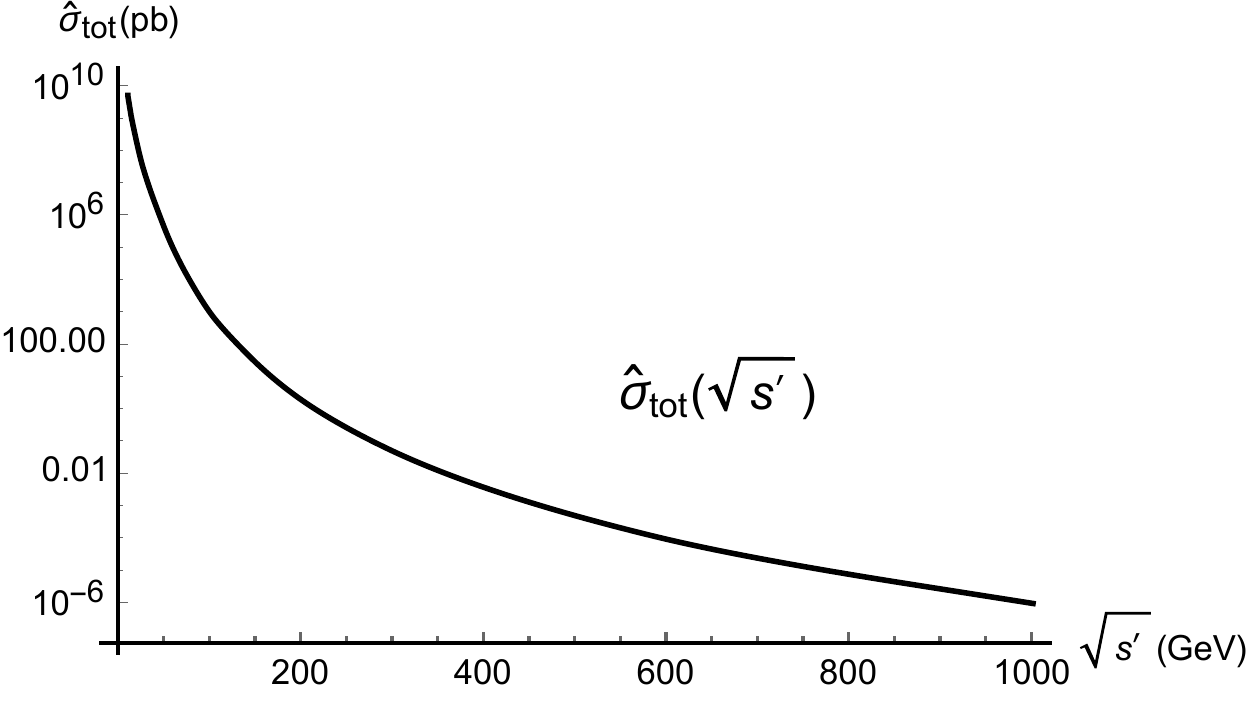}
\\
\end{tabular}
\end{center}
\vskip-.4cm
\caption{
Instanton cross-section as the function of partonic CoM energy $\sqrt{s'}$. The plot on the left shows $\hat\sigma_{\rm tot}^{\rm inst}$
for eight $q\bar{q}$ pairs in the final state (in red) and for ten $q\bar{q}$ pairs (in blue). The sum of these two contributions to the total cross-section
is shown on the right plot. The mean number of final state gluons varies with energy and can be read from the right plot in Fig.~\ref{fig:alpng}.}
\label{fig:sigmaC}
\end{figure}

The factor ${\cal K}_{\rm ferm}$ appearing in \eqref{eq:op_Pfinal} is the contribution  of $2N_f$ fermion zero modes for the light quark flavours.
Specifically, for the instanton to be able to probe $N_f=5$ fermion flavours, it is required that $m_5 < 1/\rho$ where $\rho$ is the characteristic instanton size
determined by the saddle-point for a given $\sqrt{s'}$ and $m_5$ is the mass of the $b$-quark. In this case,
 to compute the total partonic cross-section for producing
$N_f=5$ quark-anti-quark pairs in the final state we use the formula \eqref{eq:Kferm},
\begin{equation}
(5\times q_R\bar{q}_L) : \quad {\cal K}_{\rm ferm}\,=\, (\omega_{\rm\,  ferm})^{10}\,, 
\label{eq:Kferm5}
\end{equation}
with
\begin{equation} 
\omega_{\rm\,  ferm} \,= \,  
\frac{3\pi}{8}\frac{1}{z^{3/2}}\,{}_2F_{1} \left(\frac{3}{2},\frac{3}{2};4;1-\frac{1}{z^2}\right)
\,, \quad  
z\,=\, \frac{1}{2} (2+\chi^2+\chi\sqrt{4+\chi^2})\,.
\label{eq:omegaF2}
\end{equation}
But because the fermions are not strictly massless,
it is also possible to produce fewer than 5 $q_R\bar{q}_L$ pairs by saturating fermion zero modes with the fermion mass. In this case we have,
\begin{equation}
(4\times q_R\bar{q}_L) : \quad {\cal K}_{\rm ferm}\,=\, (m_5 \rho)^2 (\omega_{\rm\,  ferm})^{8} \,=\, (m_5 u/\sqrt{s'})^2 (\omega_{\rm\,  ferm})^{8}\,.
\label{eq:Kferm4}
\end{equation}
This formula applies in the regime $0 < m_5 \rho \lesssim 1$. When $m_5 \rho >1$, the instanton cannot resolve the fifth quark and one than uses 
$ {\cal K}_{\rm ferm}\,=\,(\omega_{\rm\,  ferm})^{8}$.

\medskip

\noindent In Fig.~\ref{fig:sigma} we plot the instanton production cross-section $\hat\sigma_{\rm tot}^{\rm inst}$ computed in 
\eqref{eq:sigmafin}-\eqref{eq:op_Pfinal} as a function of $\sqrt{s'}$ in picobarns for
producing $n_g$ gluons and $N_f$ quark-anti-quark pairs in the final state. The plot on the left is for $N_f=4$ and the plot on the right is for
$N_f=5$.

\begin{table}[]
\begin{center}
\begin{tabular}{|r||r|r|r||r|}
\hline
$\sqrt{s'}$ [GeV] & $1/\rho$ [GeV] & $\alpha_S(1/\rho)$ & $\langle n_g\rangle$ & $\hat\sigma$ [pb]\\
\hline\hline
10.7   &   0.99 & 0.416 & 4.59 & $4.922 \cdot 10^9$\\ 
11.4   &   1.04 & 0.405 & 4.68 & $3.652\cdot 10^9$\\ 
13.4   &   1.16 & 0.382 & 4.90 & $1.671 \cdot 10^9$\\ 
15.7   &   1.31 & 0.360 & 5.13 & $728.9 \cdot 10^6$\\
22.9   &   1.76 & 0.315 & 5.44 & $85.94 \cdot 10^6$\\ 
29.7   &   2.12 & 0.293 & 6.02 & $17.25 \cdot 10^6$\\ 
40.8   &   2.72 & 0.267 & 6.47 & $2.121 \cdot 10^6$\\
56.1   &   3.50 & 0.245 & 6.92 & $229.0 \cdot 10^3$\\
61.8   &   3.64 & 0.223 & 7.28 & $72.97 \cdot 10^3$\\ 
89.6   &   4.98 & 0.206 & 7.67 & $2.733 \cdot 10^3$\\ 
118.0  &   6.21 & 0.195 & 8.25 & $235.4\qquad$\\
174.4  &   8.72 & 0.180 & 8.60 & $6.720\qquad$\\ 
246.9  &  11.76 & 0.169 & 9.04 & $0.284 \qquad$\\
349.9  &  15.90 & 0.159 & 9.49 & $0.012\qquad $\\ 
496.3  &  21.58 & 0.150 & 9.93 & $5.112 \cdot 10^{-4}$\\
704.8  &  29.37 & 0.142 & 10.37 & $21.65\cdot 10^{-6}$\\ 
1001.8 &  40.07 & 0.135 & 10.81 & $0.9017\cdot 10^{-6}$\\ 
1425.6 &  54.83 & 0.128 & 11.26 & $36.45\cdot 10^{-9}$\\
2030.6 &  75.21 & 0.122 & 11.70 & $1.419\cdot 10^{-9}$\\
2895.5 &  103.4 & 0.117 & 12.14 & $52.07\cdot 10^{-12}$\\
\hline
\end{tabular}
\end{center}
\vskip-.4cm
\caption{
Data points for the inverse instanton radius, $1/\rho$, a leading-order value of 
$\alpha_s$, 
the expected number of gluons, $\langle n_g\rangle$ and the
partonic instanton cross-sections $\hat\sigma(s')$ of Eqs.~\eqref{eq:sigmafin}-\eqref{eq:op_Pfinal} in the range of $10$ GeV  -- $3$ TeV.}
\label{fig:table2}
\end{table}

\medskip
A selection of our theory prediction data-points for parton-level instanton processes is presented in Table~\ref{fig:table2} for
a broad partonic energy range $10\, {\rm GeV} < \sqrt{s'} < 2\, {\rm TeV}$.

\medskip
\section{\label{Sec:Sherpa}Implementation in the \Sherpa Event Generator}
\medskip

Modelling instanton-induced processes is achieved by multiplying the partonic cross section $\hat\sigma(s')$ with parton distribution functions and integrating over the initial state,
\begin{equation}
    \sigma_{I}(s'>s'_{\rm min}) = 
    \int\limits_{s'_{\rm min}}^{s_{pp}} \mathrm{d}x_1\mathrm{d}x_2
    \sum\limits_{i,j} f_{i}(x_1,\,\mu_F)f_{j}(x_2,\mu_F)
    \hat\sigma_{ij\to I}(s'=x_1x_2s_{pp})\,,
    \label{eq:hadxsec}
\end{equation}
where $s'_{\rm min}$ is the minimal invariant mass squared of the produced system and $s_{pp}$ is the CoM energy squared of the colliding protons.  Note that below we present details of the simulation for the purely gluon-initiated process, the extension to also include quarks in the initial state is trivial.

In \Sherpa~\cite{Gleisberg:2008ta,Bothmann:2019yzt} the partonic instanton production cross section is obtained as functions of the partonic CoM energy squared $s'$ through linear interpolation from the values listed in Table \ref{fig:table2}, that have been hard-coded.  In the code, we allow the user to specify the lower and upper limit of the systems squared mass $s'$, and we also provide the possibility to multiply the partonic cross sections with an additional, user-defined factor to allow for some systematic checks.  
\begin{table}[]
\medskip
    \centering
    \begin{tabular}{|c||r|r|r|r|r|}
    \hline
    $\sqrt{s'_{\rm min}}$ [GeV] &   20      & 50          & 100     & 200     & 500\\
    \hline\hline
    $\sigma_{pp\to I}$          &  6.32 mb & 40.82 $\mu$b & 79.95 nb & 105.4 pb & 3.54 fb\\
    \hline
    \end{tabular}
    \caption{Hadronic cross sections for instanton production through initial gluons,
        at the 13 TeV LHC, using the NNPDF3.1 NNLO set with $\alpha_s(M_Z)=0.118$~\cite{Ball:2017nwa}.  }
    \label{tab:hadronic_xsecs}
\end{table}

Choosing the $s'$ according to the distribution emerging from Eq.\ (\ref{eq:hadxsec}) and the rapidity of the system $\hat y$ flat in its allowed region fixes the overall kinematics of the system emerging in the final state, and the selected $s'$ also fixes the default factorization scale $\mu_F = 1/\rho$ and the mean number of gluons, $\langle n_g\rangle$.  In \Sherpa we also provide an alternative choice for the factorization scale, namely $\mu_F=\sqrt{s'}$.  Hadronic cross sections for different choices of $\hat s'_{\rm min}$, with the default choice of $\mu_F=1/\rho$, and using the NNPDF3.1 NNLO distribution~\cite{Ball:2017nwa}, are listed in Table~\ref{tab:hadronic_xsecs}.\footnote{
    Note, that for $\sqrt{s'}$ below about 20 GeV the scale $1/\rho$ falls below the minimal $\mu_F$ for which the PDF has support. For such low values one can
    use $\mu_F = Q_{\rm min} = 1.65$ GeV, the minimal scale for which there is
    support.} 
The large hadronic cross section of about 6~mb for $\sqrt{s'_{\rm min}} = 20$~GeV 
-- about 5\% of the total proton-proton cross section -- and the strong increase with smaller minimal instanton masses suggests that for even smaller mass/energy ranges the cross section will saturate the $pp$ cross section and therefore becomes untrustworthy.  This implies that to regularise the cross section for smaller masses additional effects have to start playing a significant role.

To specify the particle content of the final state, we add quark--anti-quark pairs $q\bar q$, subject to two constraints:
\begin{enumerate}
    \item the mass of the quark $m_q$ has to be smaller than a kinematics dependent
    threshold $\mu_q$, $m_q<\mu_q$.  In the simulation we offer two options, namely $\mu_q = E'=\sqrt{s'}$ (the default we use in the following), and $\mu_q = 1/\rho$.
    \item we also demand that the combined mass of all pair-produced quarks is smaller
    then $E'$ and stop adding more quark pairs once we saturated this constraint.
\end{enumerate}
After that we select the number of additional gluons $n_g$ according to a Poissonian distribution with mean $\langle n_g\rangle$, which can be modified by a user-defined multiplier (set to 1 by default).  Momenta of the outgoing particles are generated through the \Rambo algorithm~\cite{Kleiss:1985gy}.  It produces $n$ isotropically distributed momenta in their own rest-frame and characterised by an invariant mass $M=E'$.  The overall system is then boosted back from its rest frame to the lab frame.  Finally, the colours of the quarks and gluons entering and leaving the process are randomly distributed, and only subject to the condition of overall colour conservation.

In the \Sherpa simulation, the subsequent parton showers~\cite{Schumann:2007mg,Hoche:2015sya} in the initial and final state start at the scale $\mu_Q$.  It is given by evaluating the maximal transverse momentum of outgoing single partons has with respect to their colour partner(s).  After the parton showers terminate, the events can be further supplemented with the usual multi-parton interactions and the emerging partons will hadronize~\cite{Winter:2003tt}.   

\medskip
\section{\label{Sec:Experiment}Experimental Signatures}

It is well known from previous searches for QCD instantons at the HERA collider \cite{Chekanov:2003ww, H1:2016jnv} that experimental signatures of instanton-induced processes in high energy collisions are difficult to distinguish from other standard model processes.  The H1 and Zeus Collaborations at HERA expected isotropic decays in the sphaleron rest-frame into ${\cal{O}}(10)$ partons ("fire-ball"), leading to a band structure in a defined pseudo-rapidity region of the detector. Since all light quark flavours are equally present in the final state (flavour democracy), several strange mesons and baryons such as $K^\pm$ and $\Lambda$'s should be observed. In addition, the current quark defining the virtuality of the process leads to one highly energetic jet in the forward region. The discrimination of the instanton-induced contribution and their backgrounds were based on the objects in the hadronic final state, and primarily on observables constructed from the charged particles. A multivariate discrimination technique was employed by H1 to increase the sensitivity to instanton processes, leading to the strongest upper limits (\cite{Carli:1998zf}). They range between 1.5 pb and 6 pb, at 95\% confidence level, depending on the chosen kinematic domain. While this result challenges the predictions based on the lattice data of Ref. \cite{Smith:1998wt}, it is fully compatible \cite{Schremp2016p} with the expectations based on the lattice data of Ref. \cite{Hasenfratz:1998qk}. \\

For the experimental search for QCD instanton-induced processes in proton-proton collisions, we treat the final state of the instanton-process \eqref{eq:inst1} as if it was produced in a decay of a pseudo-particle with a mass above $\sqrt{s'_{\rm min}}$, cf.\ \eqref{eq:inst1}. While {\bf low instanton masses} ($\sqrt{s'}\approx 30$ GeV) will lead to few isotropic tracks with energies of a few GeV in the detector, in the regime of {\bf high instanton masses} ($\sqrt{s'}\approx 500$ GeV) we expect numerous isotropic particle-jets with energies of around or more than 20 GeV.  In the low-mass regime, we expect mainly pile-up and underlying event activities as well as low energetic hard QCD scattering of partons in proton-proton collisions as background processes. In the high-mass domain the dominant background processes will be the production of hadronically decaying top-quark pairs or $W$ bosons in association with jets as well as hard QCD scattering processes leading to multi-jet events.

In contrast to typical searches for new particles, we explicitly expect no resonance behaviour, but rather a continuous, rapidly falling spectrum of invariant masses of the instanton-produced hadronic final states, governed by Eq.~\eqref{eq:op_th3}. This implies significant challenges in the search for an evidence of instanton-induced processes: while sizeable cross-sections are expected for small instanton masses, the experimental signatures in this energy regime might be difficult to distinguish from non-perturbative QCD effects, such as underlying event activities, or, at high luminosities, the large pile-up.  Since these backgrounds can be only described by a combination of data and phenomenological model with a significant number of tunable parameters, it will be challenging to prove that discrepancies between data and those models are due to instanton processes. On the other hand, the experimental signatures of instanton-induced processes are very striking in the high energy regime; however, their cross sections are then largely suppressed and hence difficult to observe in the first place.

One possible approach to tackle these challenges is using the energy dependence of instanton processes, which is well predicted and significantly different from various other SM processes. Once finding experimental observables, which are different for instanton final states and other SM processes, their dependence on the instanton mass might be used as additional leverage. A dedicated search strategy will therefore be based on a simultaneous analysis over the full available energy regime at the LHC, investigating simultaneously several observables.

In the following, we will discuss some selected and indicative observables for the two mass ranges, $\sqrt{s'}_{\rm min}=30$ GeV and $\sqrt{s'}_{\rm min}=500$ GeV, as well as the expected background processes. Clearly, this is meant only as a  first look into possible observables and mainly serves as a motivation for future studies which will take into account the composition and impact of backgrounds in more detail. 

All background processes have been produced with the \textsc{Pythia8} \cite{Sjostrand:2007gs} event generator, using the \textsc{CT10nlo} PDF set \cite{Dulat:2015mca} and standard \textsc{Pythia8} tune settings. An overview is shown in Table \ref{tab:samples}. A typical detector response has been simulated through the \textsc{Delphes}-framework \cite{deFavereau:2013fsa} using the settings of the ATLAS experiment.

\begin{table}[thb]
\medskip
\footnotesize
\centering
\begin{tabular}{|l| c|c|c|}
\hline
{\bf Process}								& {\bf Generator}	& {\bf Main Generator Setting}					& {\bf \# Events}	\\
\hline
QCD-instanton (low-mass regime)				& \textsc{Sherpa}	& INSTANTON\_MIN\_MASS: 30.										& 10,000			\\
QCD-instanton (high-mass regime)				& \textsc{Sherpa}	& INSTANTON\_MIN\_MASS: 500.										& 1,000			\\
\hline
Soft-QCD									& \textsc{Pythia8}	& \textsc{SoftQCD:all = on}					& 100,000			\\
\hline
$qq\rightarrow X, qg\rightarrow X, gg\rightarrow X$ 	& \textsc{Pythia8}	& \textsc{HardQCD:all = on}					& 100,000	\\
(Hard-QCD, low energy)						&				& \textsc{PhaseSpace:pTHatMin = 5.}			&		\\
\hline
$qq\rightarrow X, qg\rightarrow X, gg\rightarrow X$	& \textsc{Pythia8}	& \textsc{HardQCDAll=on}					& 100,000	\\
(Hard-QCD, high energy)						&				& \textsc{PhaseSpace:pTHatMin = 100.}			&		\\
\hline
$W\rightarrow q\bar q+X$						& \textsc{Pythia8}	& \textsc{WeakSingleBoson:ffbar2W = on}			& 100,000					\\
\hline
$t\bar t\rightarrow b q\bar q+\bar b q\bar q+X$		& \textsc{Pythia8}	& \textsc{Top:all = on}						& 100,000			\\
\hline
\end{tabular}
\caption{\label{tab:samples} Overview of MC samples used to study observables that allow to discriminate signal and potential background processes.}
\end{table}

In each event, we first sum over all reconstructed 4-vectors of charged particles tracks with transverse momenta above 500 MeV and particle jets with transverse energies above 20 GeV. Particle jets are reconstructed using an anti-k$_T$ algorithm with a cone-size of 0.4. The resulting invariant mass can be taken as proxy for the instanton mass, denoted as $M^{\rm reco}_I$ in the following. The relevant observables for events with $20<M^{\rm reco}_I<30$ GeV (low-mass) will be based on reconstructed tracks, while they will be based on reconstructed jets for events with $320<M^{\rm reco}_I<480$ GeV (high-mass). These limited kinematic regions lead to a nearly constant $M^{\rm reco}_I$ spectra, hence the resulting distributions can be compared on an equal footing. It should be also noted, that $M^{\rm reco}_I$ is typically smaller than $\sqrt{s'}$ since not all final state objects get reconstructed. 

\begin{figure}[htbp]
\begin{center}
\includegraphics[width=0.495\textwidth]{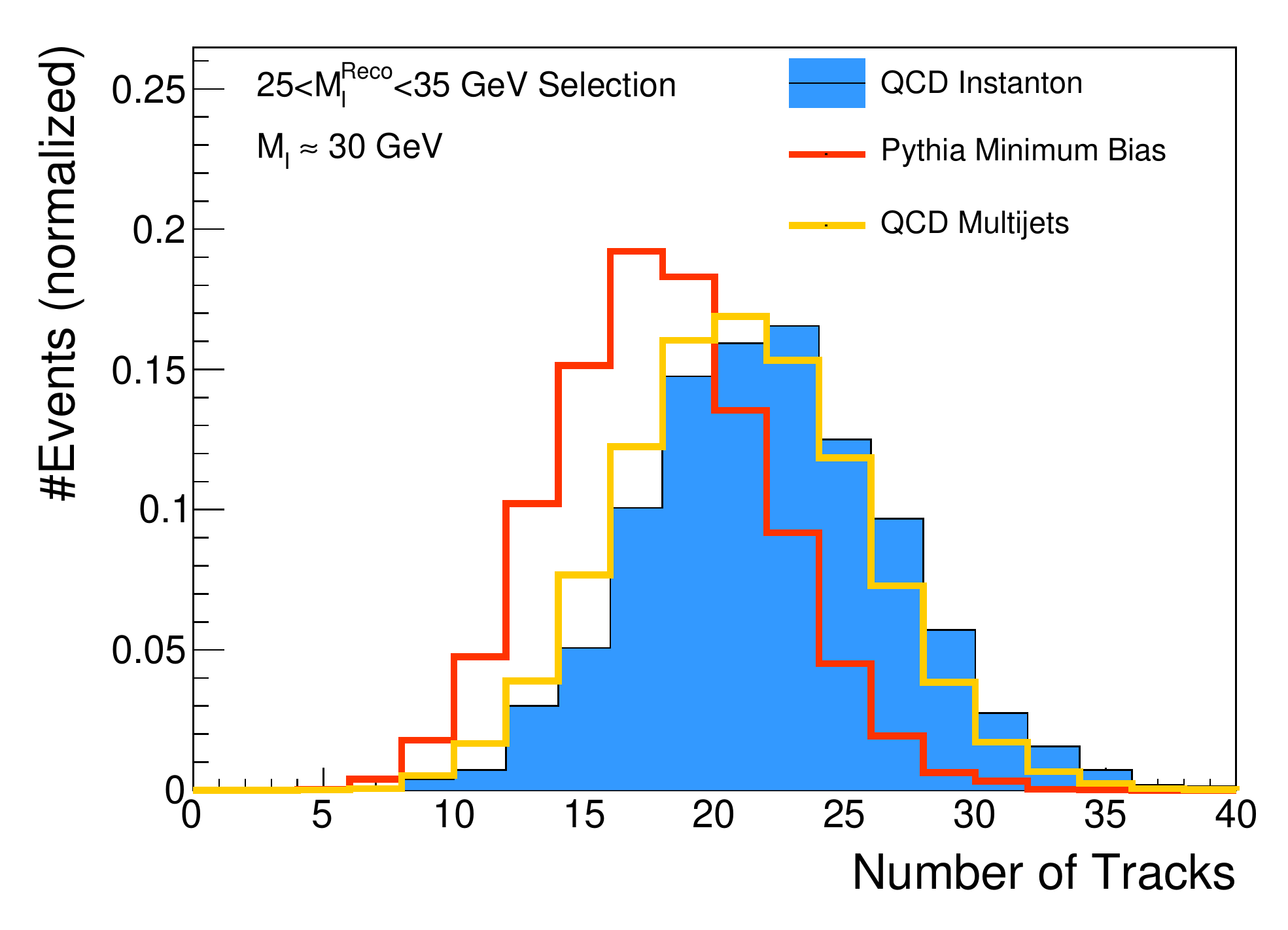}
\includegraphics[width=0.495\textwidth]{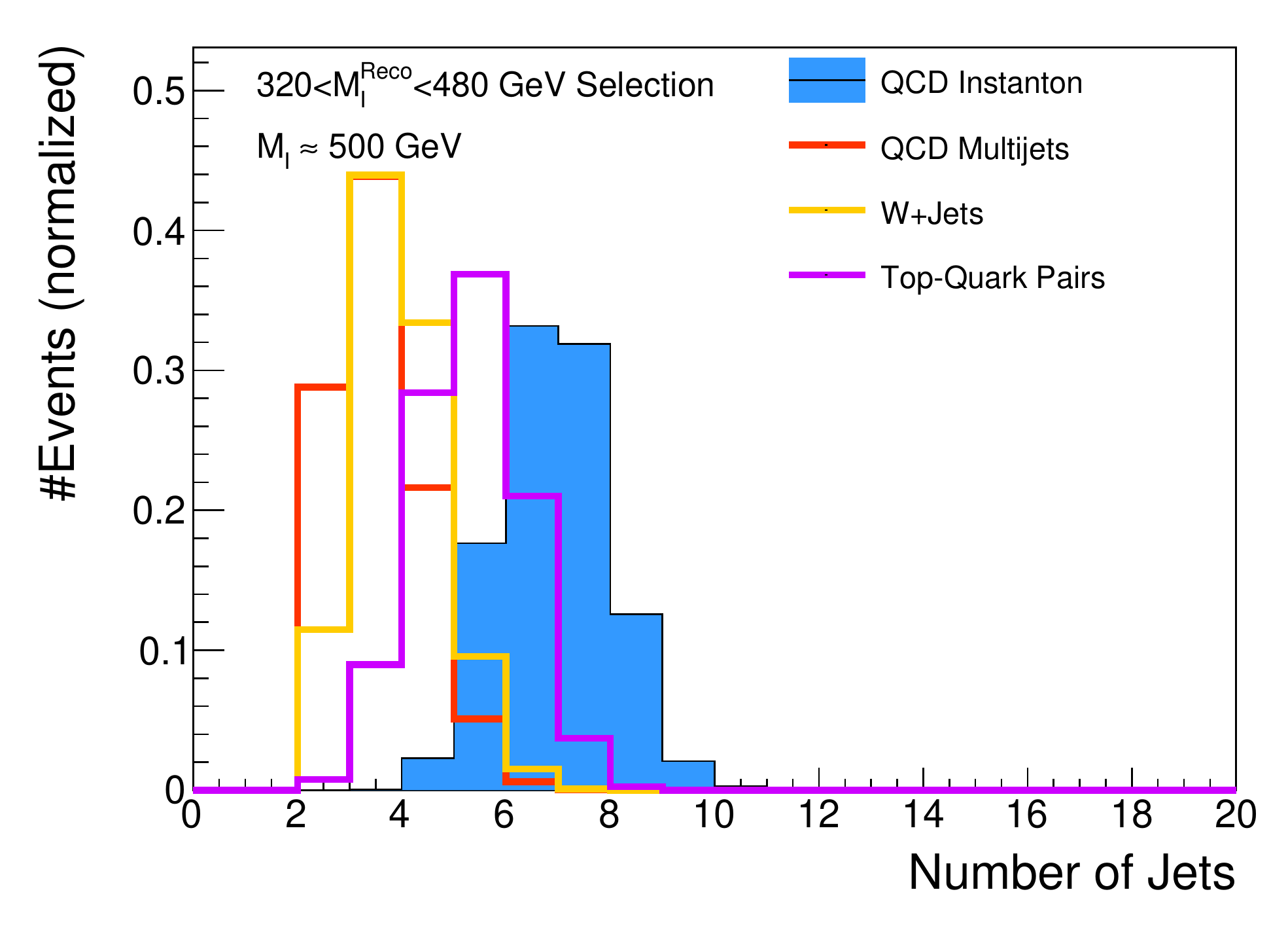}
\caption{\label{fig:NTracks}Normalized distribution of the number of reconstructed tracks for events with $25<M^{\rm reco}_I<35$ GeV (left) and of reconstructed particle-jets for events with $320<M^{\rm reco}_I<480$ GeV (right). Beside the signal processes, the expected distributions of the background processes are shown (see Table \ref{tab:samples}).}
\end{center}
\end{figure}

A first observable, potentially sensitive to QCD instanton decays, is the number of reconstructed tracks and jets for a given range of $M^{\rm reco}_I$. The relevant distributions for the expected signal and relevant background processes are shown in Figure~\ref{fig:NTracks} for the low and high mass case. Note that all distributions are normalised to unity, i.e.\ only the expected shapes are compared and cross sections are not accounted for.  This, in fact is a sensible approach, because, as discussed above, the calculation of the instanton cross section and its result are subject to a number of assumptions and approximations.  In the low-mass case, we observe on average more tracks for the signal processes, while this effect is even more pronounced for the number of reconstructed jets in the high-mass case.

\begin{figure}[htbp]
\begin{center}
\includegraphics[width=0.495\textwidth]{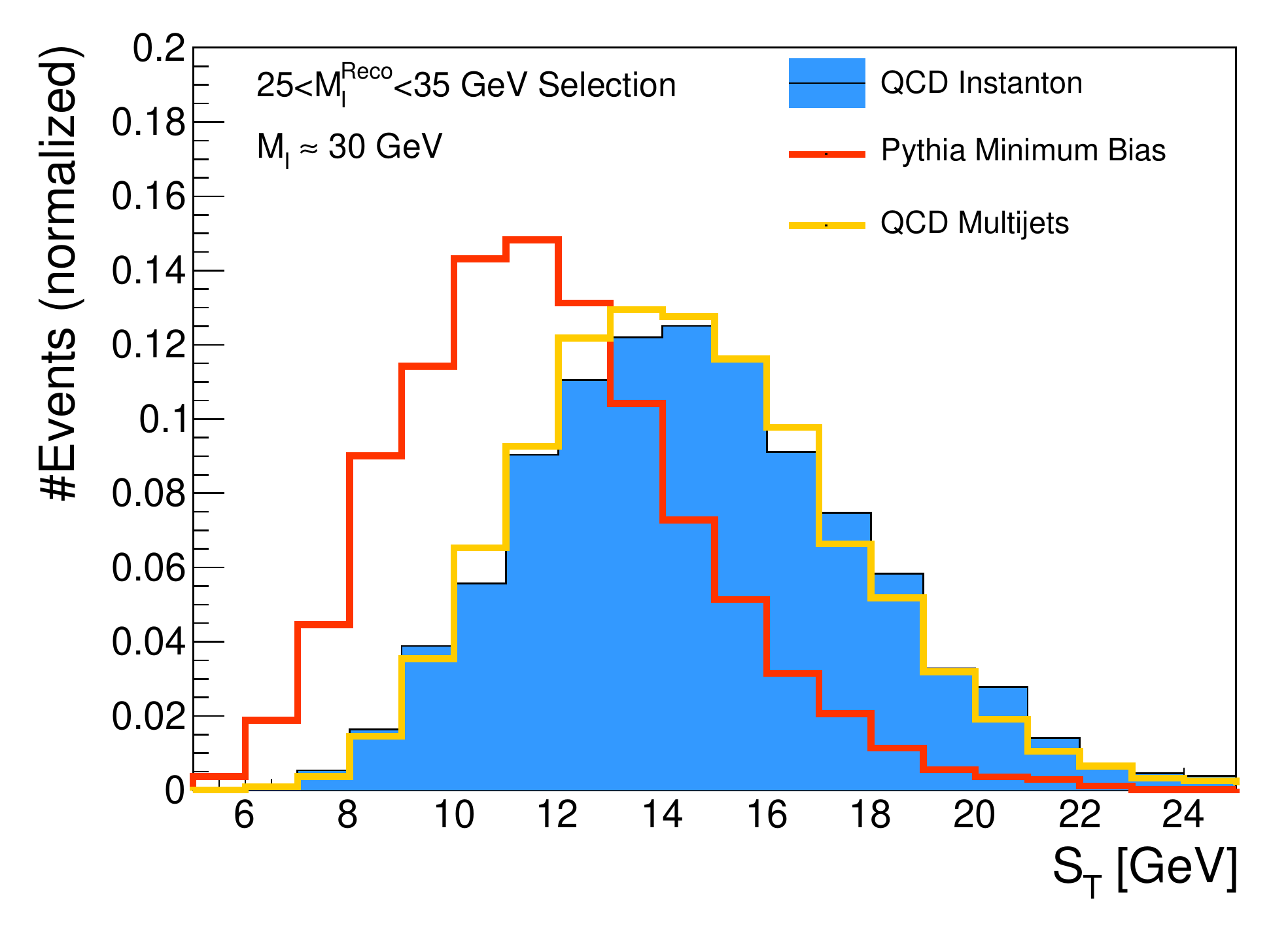}
\includegraphics[width=0.495\textwidth]{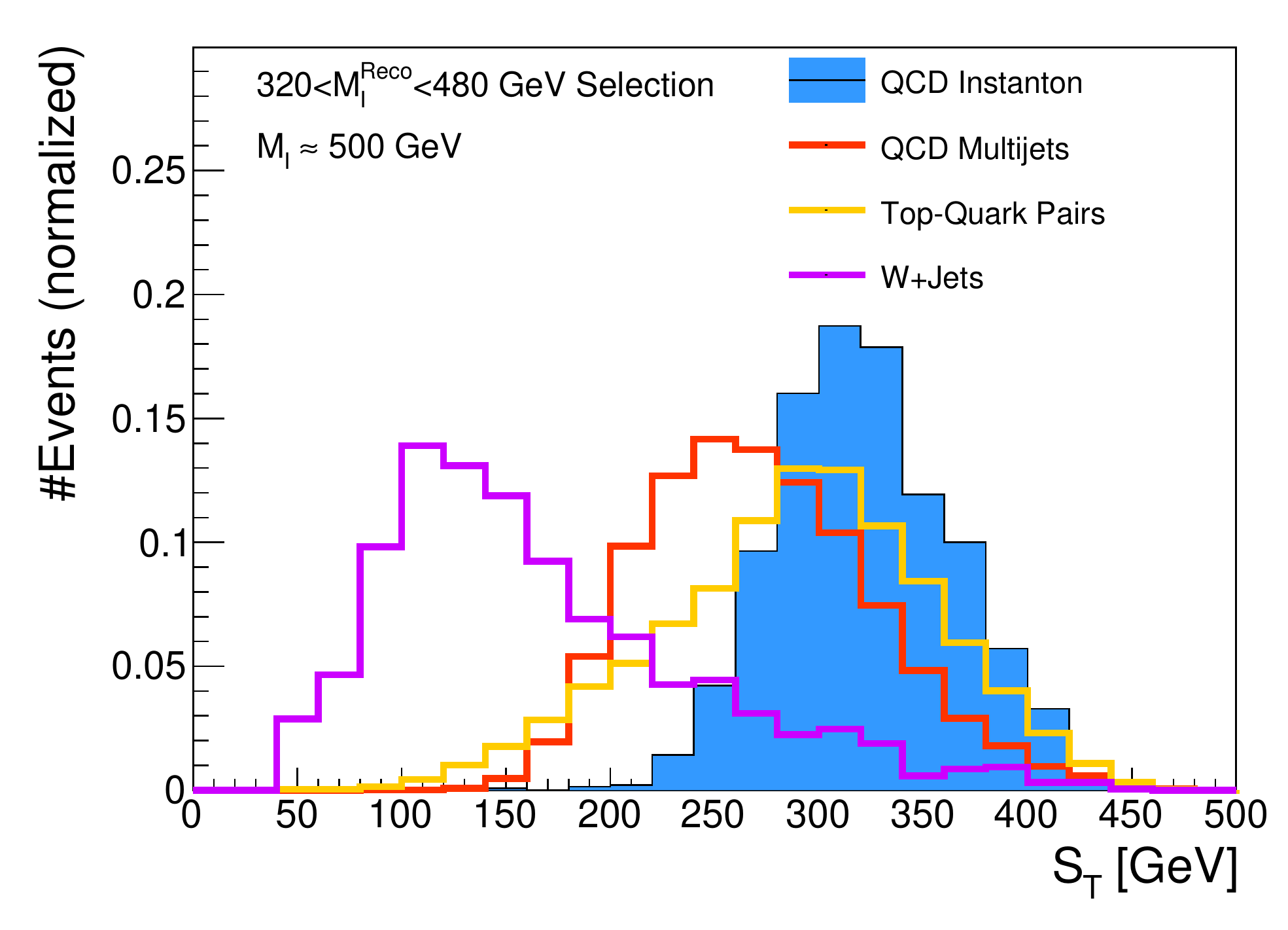}
\caption{\label{fig:STs}Normalized $S_T$ distributions of reconstructed tracks for events with $20<M^{\rm reco}_I<30$ GeV (left) and of reconstructed particle-jets for events with $320<M^{\rm reco}_I<480$ GeV. Beside the signal processes, the expected distributions of the background processes are shown (see Table \ref{tab:samples}).}
\end{center}
\end{figure}

A similar behaviour is seen for the scalar sum of all transverse momenta of reconstructed tracks and jets, i.e.\ $S_T=\sum_i p_T^i $, shown in 
Figure~\ref{fig:STs}. The scalar sum is expected to be on average higher for the signal compared to the background processes and the difference becomes more significant for the high-mass case.

Since it is expected that the instanton decay results in an isotropic final state distribution of particles, it is worth to define the average angle between all reconstructed objects in the transverse-plane of the detector, i.e.\
\begin{equation}
\langle\Delta \phi\rangle \,=\, \frac{1}{N} \sum_{i,j,i \neq j} \Delta \phi (i,j).
\end{equation}

\begin{figure}[htbp]
\begin{center}
\includegraphics[width=0.495\textwidth]{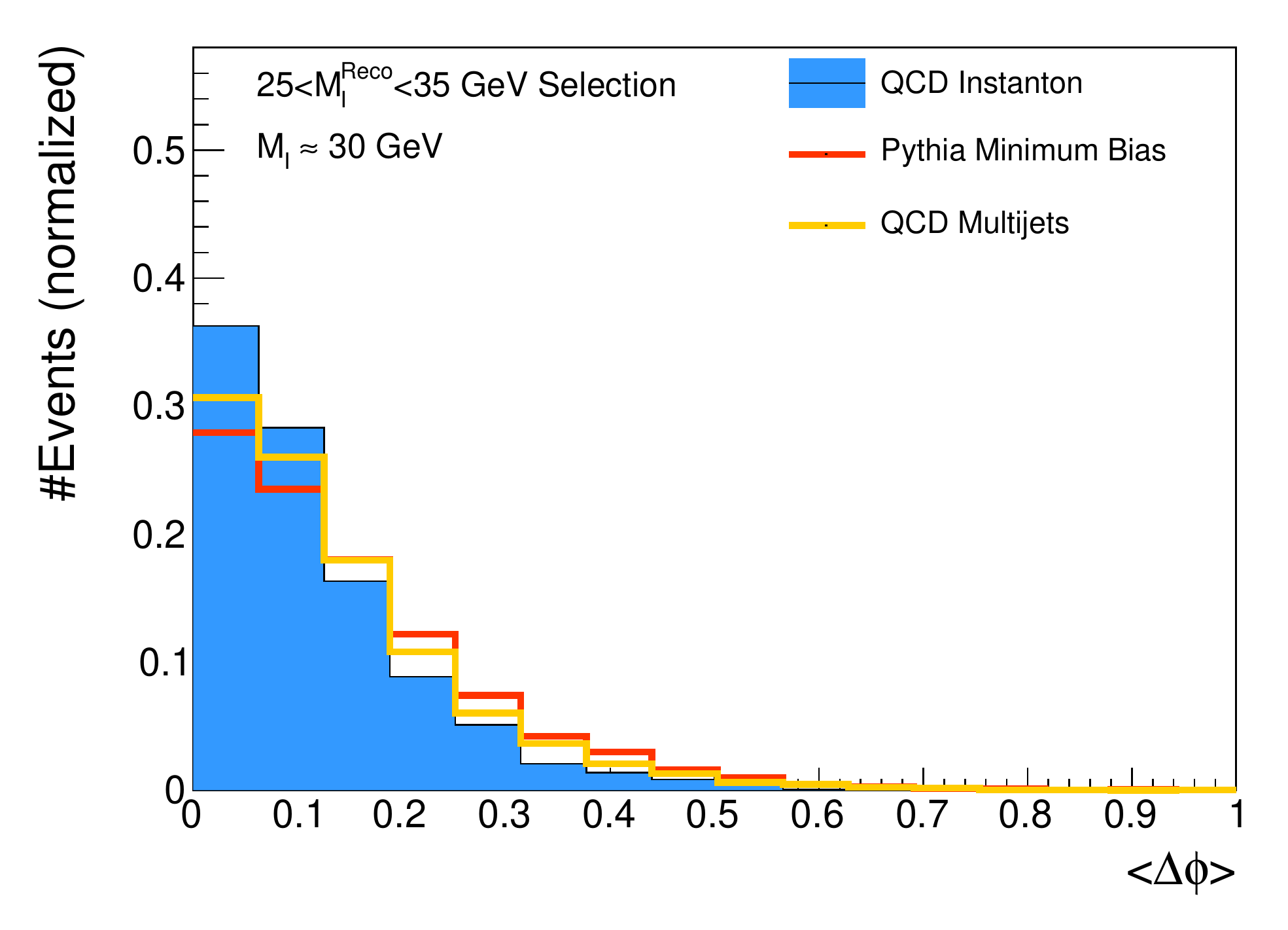}
\includegraphics[width=0.495\textwidth]{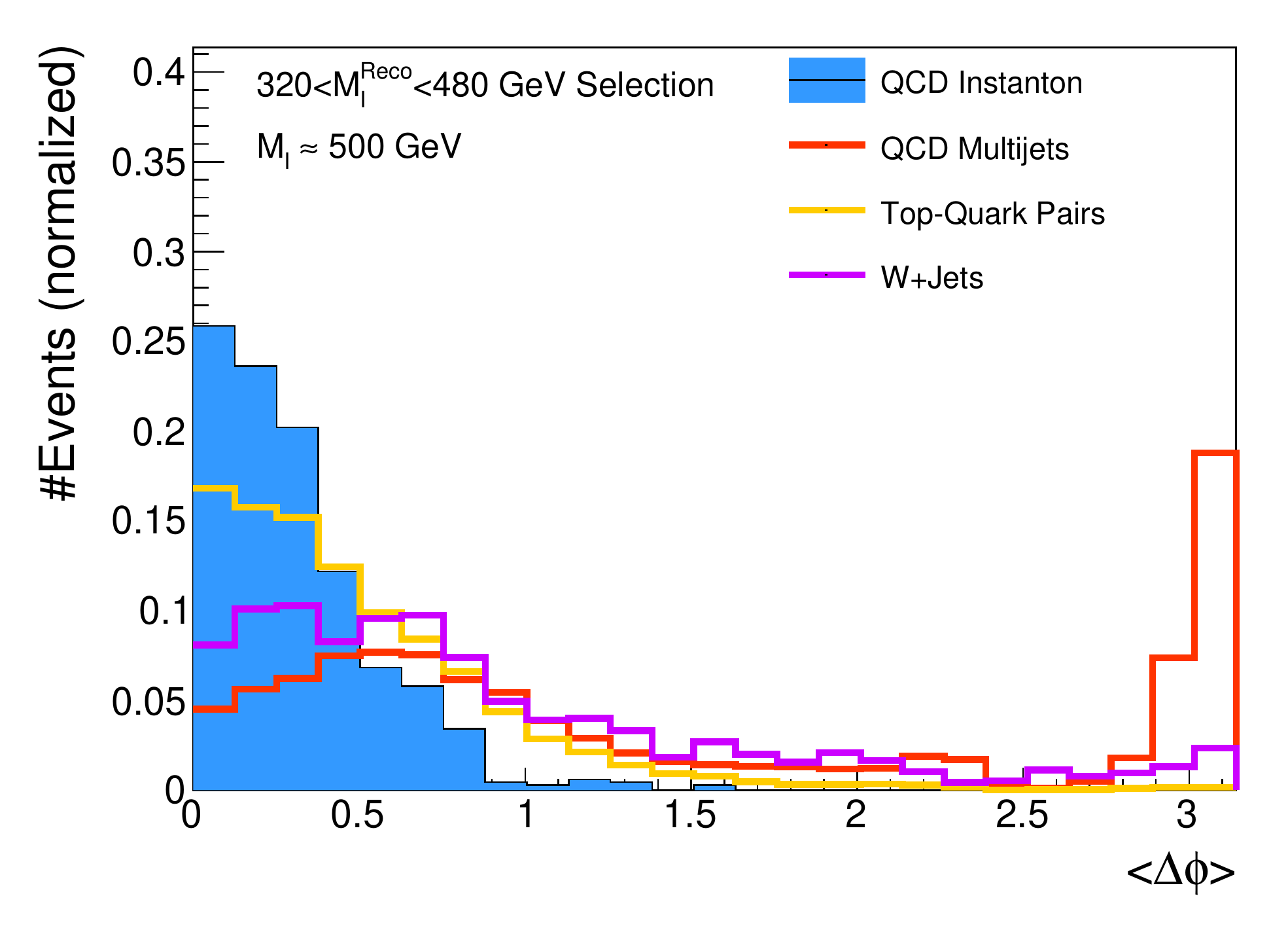}
\caption{\label{fig:AveragePhi}Normalized distributions for average angles between reconstructed tracks for events with $25<M^{\rm reco}_I<35$ GeV (left) and between reconstructed particle-jets for events with $320<M^{\rm reco}_I<480$ GeV (right). Beside the signal processes, the expected distributions of the background processes are shown (see Table \ref{tab:samples}).}
\end{center}
\end{figure}

The corresponding distributions of signal and background processes are shown in Fig.~\ref{fig:AveragePhi} for both cases. As expected, we observe on average a smaller value of $\langle\Delta \phi\rangle$ for the instanton decay processes. 

\begin{figure}[htbp]
\begin{center}
\includegraphics[width=0.495\textwidth]{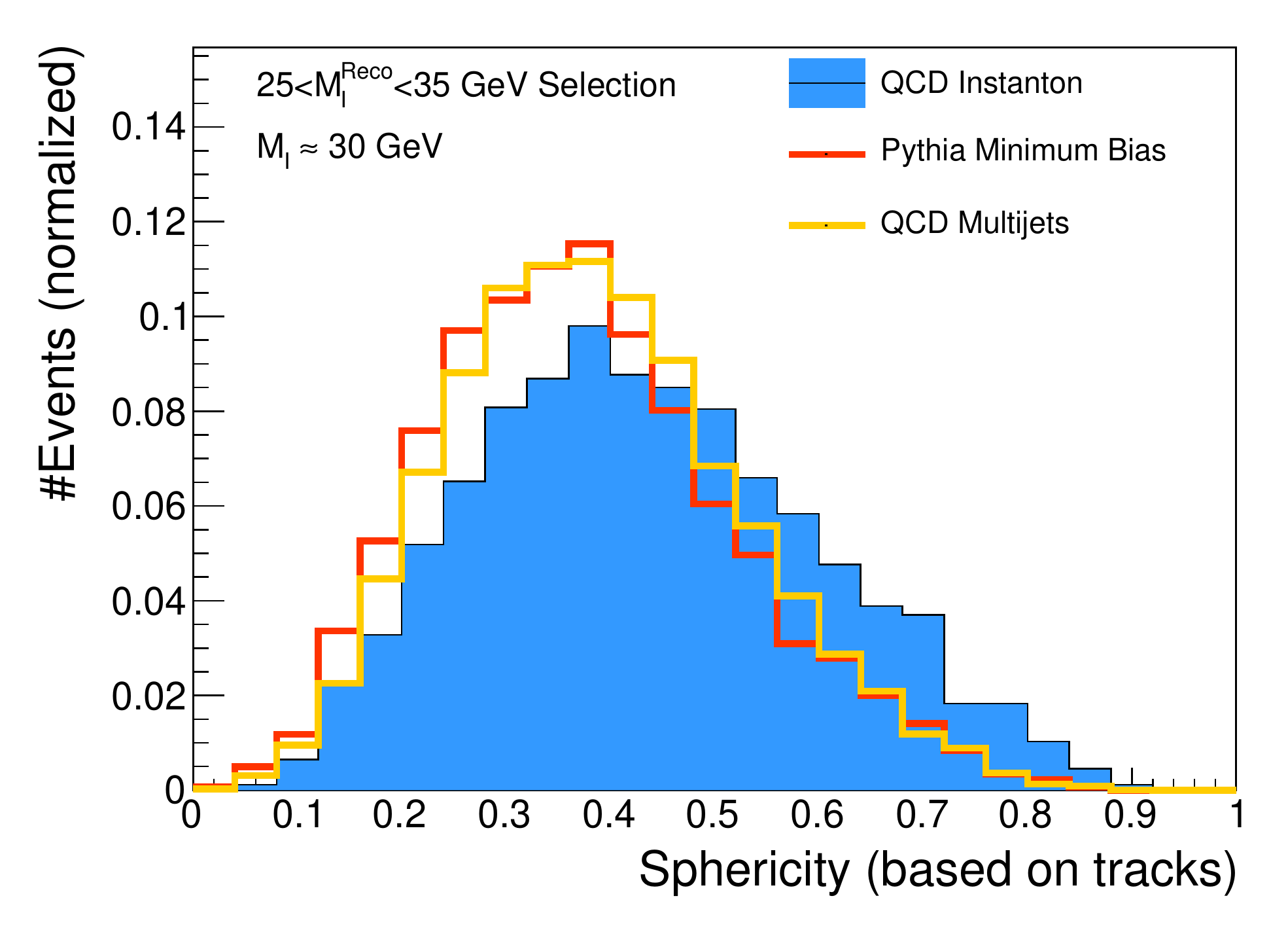}
\includegraphics[width=0.495\textwidth]{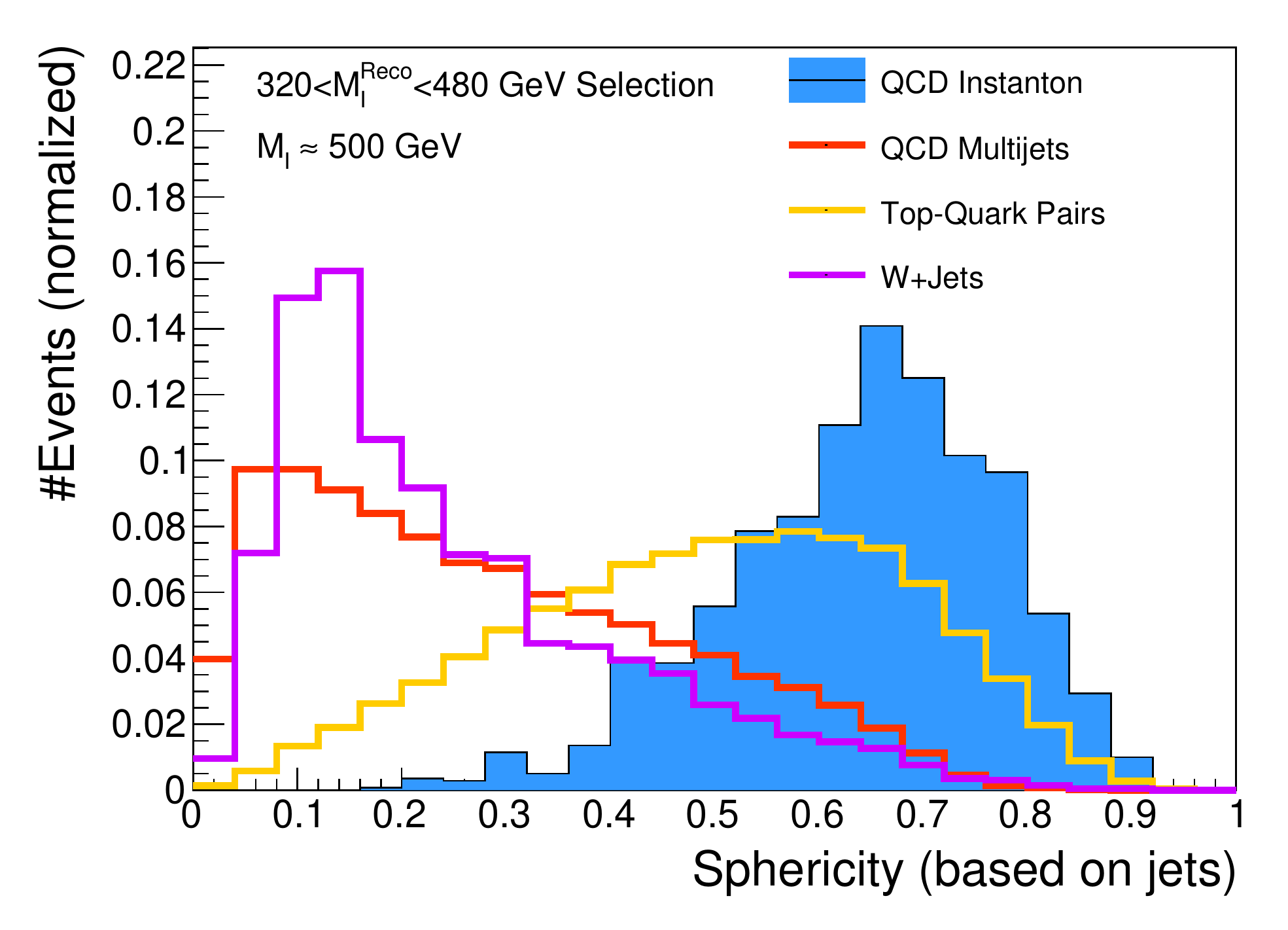}
\caption{\label{fig:Spherity}Normalized spherity distributions of reconstructed tracks for events with $25<M^{\rm reco}_I<35$ GeV (left) and of reconstructed particle-jets for events with $320<M^{\rm reco}_I<480$ GeV (right). Beside the signal processes, the expected distributions of the background processes are shown (see Table \ref{tab:samples}).}
\end{center}
\end{figure}

An alternative observable that targets the isotropy of an event, is called sphericity and is defined via the tensor $S$,
\begin{equation}
	S^{\rm \alpha \beta} \,=\,  \frac{\displaystyle\sum_{i} p^{\rm \alpha}_{i} p^{\rm \beta}_{i}}{\displaystyle\sum_{i} |{\vec{p}_{i}}|^{2}}~\mbox{,}\nonumber
\end{equation}
where the indices denote the $x$, $y$, and $z$ components of the momentum of the particle $i$. The sphericity of the event is then constructed using the two smallest eigenvalues of this tensor, 
$\lambda_2$ and $\lambda_3$, i.e.\ $S=\frac{3}{2}({\lambda_2 + \lambda_3})$ and takes values between 0 and 1. A fully balanced dijet events leads to a spherity of $S=0$, while a fully isotropic event has a sphericity of $S=1$. Figure \ref{fig:Spherity} shows the sphericity distributions for the low and high mass case for the signal and the relevant background processes. Here we observe significant difference for the low-mass and high-mass case.

These observables give a first indication of how a dedicated QCD instanton search can be developed at the LHC. However, it should be stressed that the presented studies only give a first glimpse on the experimental features of QCD instanton processes at the LHC and a details for this dedicated search strategy are still to be developed. Certainly there are many more interesting observables, such as further event shape variables, variables based on flavour-tagging or direct particle identification. Most background processes for large instanton masses ($>100$) GeV can be estimated in data-driven ways, for example for $W/Z+jets$ and $t \bar t$ by using their leptonic decay channels. The situation is somewhat more complicated for the low-mass regime, as most background processes are inherent QCD phenomena which can hardly be selected without possible contributions from instanton decays. However, a combination of all accessible observables as well as their predicted dependence on the reconstructed instanton mass might allow for a first observation at the LHC and therefore provide a first experimental proof of the non-trivial vacuum structure of non-abelian gauge theories.

\section{\label{Sec:Conclusion}Conclusions}

This paper provides a detailed calculation of non-perturbative contributions to high-energy scattering processes generated by QCD instantons.
We develop and pursue a semiclassical instanton approach that accounts for quantum corrections arising 
from both initial and final-state interactions in the instanton background 
combining the methods of \cite{Khoze:1991mx} and \cite{Mueller:1990ed}.  These quantum effects provide a dynamical cut-off of QCD instantons with
 large sizes. Our results suggest that small-size instantons can be effectively produced 
and probed at colliders.

The corresponding \Sherpa implementation of instanton production, based on this calculation will be made publicly available in the forthcoming release of version 3.0.  We used it to study the effect of instantons on observable quantities at the LHC.
Our preliminary experimental studies show that QCD instantons provide novel and interesting search grounds for distinctive non-perturbative effects 
in QFT in high-energy collisions.

\section*{Acknowledgements}

We would like to thank A.~Ringwald and M.~Spannowsky for useful discussions. MS would like to thank in addition the Alexander von Humboldt foundation for the award of the Feodor Lynen research scholarship for this research topic at UCL and his host J. Butterworth.
Research of VVK and FK is partially supported by the STFC consolidated grant ST/T001011/1.  
FK acknowledges funding from the European Unions Horizon 2020 research and innovation programme as part of the Marie Sklodowska-Curie Innovative Training Network MCnetITN3 (grant agreement no. 722104).  VVK also thanks the Munich Institute for Astro- and Particle Physics (MIAPP) of the DFG Excellence Cluster Origins, where part of this work was done.

\bibliographystyle{unsrt}
\bibliography{main}

\clearpage
\appendix
\include{variables}

\end{document}